\documentclass[10pt]{article}
\usepackage{graphicx}
\usepackage{rotating}
\usepackage{color}
\renewcommand{\baselinestretch}{.97}
\begin{document}
\begin{center}
{\Large \bf Disentangling random thermal motion of particles and
collective expansion of source from transverse momentum spectra in
high energy collisions}

\vskip.5cm

Hua-Rong Wei$^{a}$, Fu-Hu Liu$^{a,}${\footnote{E-mail:
fuhuliu@163.com; fuhuliu@sxu.edu.cn}}, and Roy A.
Lacey$^{b,}${\footnote{E-mail: Roy.Lacey@Stonybrook.edu}}

{\small\it $^a$Institute of Theoretical Physics, Shanxi
University, Taiyuan, Shanxi 030006, China

$^b$Departments of Chemistry \& Physics, Stony Brook University,
Stony Brook, NY 11794, USA}
\end{center}

\vskip.25cm

{\bf Abstract:} In the framework of a multisource thermal model,
we describe experimental results of the transverse momentum
spectra of final-state light flavour particles produced in
gold-gold (Au-Au), copper-copper (Cu-Cu), lead-lead (Pb-Pb),
proton-lead ($p$-Pb), and proton-proton ($p$-$p$) collisions at
various energies, measured by the PHENIX, STAR, ALICE, and CMS
Collaborations, by using the Tsallis-standard (Tsallis form of
Fermi-Dirac or Bose-Einstein), Tsallis, and two- or
three-component standard distributions which can be in fact
regarded as different types of ``thermometers" or ``thermometric
scales" and ``speedometers". A central parameter in the three
distributions is the effective temperature which contains
information on the kinetic freeze-out temperature of the emitting
source and reflects the effects of random thermal motion of
particles as well as collective expansion of the source. To
disentangle both effects, we extract the kinetic freeze-out
temperature from the intercept of the effective temperature ($T$)
curve as a function of particle's rest mass ($m_0$) when plotting
$T$ versus $m_0$, and the mean transverse flow velocity from the
slope of the mean transverse momentum ($\langle p_T \rangle$)
curve as a function of mean moving mass ($\overline{m}$) when
plotting $\langle p_T \rangle$ versus $\overline{m}$.
\\

\\

\\

{\section{Introduction}}

Comparing with fixed target experiments, the Relativistic Heavy
Ion Collider (RHIC) in the USA and the Large Hadron Collider (LHC)
in Switzerland attract more studies due to their exhibitions on
the evolution process of interacting system in collisions at
higher energies. The experiments at RHIC and LHC are more complex
and difficult for some limited conditions, which render very
limited measurable quantities. It is expected that more
interacting information can be extracted from these limited
measurable quantities. For example, by analyzing the transverse
momentum spectra of final-state particles, one can obtain the
kinetic freeze-out temperature of interacting systems (emission
sources) and the transverse flow velocity of produced particles.
These two quantities reflect the excited degree of emission
sources and hydrodynamic expansion picture of interacting systems
at the stage of kinetic freeze-out, when hadrons are no longer
interactive and their momenta do not change [1].

In high energy collisions, one can use different distribution laws
to describe the transverse momentum spectra of final-state
particles. In the framework of a multisource thermal or
statistical model, we can use the standard (Boltzmann,
Fermi-Dirac, or Bose-Einstein) distribution [2], the
multi-component standard distribution, the Tsallis distribution
(statistics), the Tsallis-standard (Tsallis forms of the standard)
distribution [3], the Erlang distribution, the multi-component
Erlang distribution [4], the L\'evy distribution [5, 6], the
blast-wave function [7], the power law, and so forth, to describe
the particle transverse momentum spectrum contributed by a given
emission source. From the distributions mentioned above, one can
directly extract the effective temperature, which is actually not
the real temperature (kinetic freeze-out temperature) of the
emission source. The real temperature of the emission source
should reflect purely thermal motion of particles in the source,
while the effective temperature extracted from the transverse
momentum spectrum includes both thermal motion and flow effect of
particles, and is greater than the real temperature. Only if the
effect of flow is excluded in the extraction of source
temperature, can we obtain the kinetic freeze-out temperature.

In this paper, within the framework of the multisource thermal
model [8--10], we use the Tsallis-standard distribution, Tsallis
distribution, and two- or three-component standard distribution
[3] to describe the transverse momentum spectra of final-state
light flavour particles produced in gold-gold (Au-Au),
copper-copper (Cu-Cu), lead-lead (Pb-Pb), proton-lead ($p$-Pb),
and proton-proton ($p$-$p$) collisions with different centrality
intervals over a center-of-mass energy ($\sqrt{s_{NN}}$) range
from 0.2 to 7 TeV. Although we can obtain the analytical
expressions for the mentioned distributions, the Monte Carlo
method is used to see the statistical fluctuations in the process
of calculation. The results from the Monte Carlo method are
compared with the experimental data of the PHENIX [11], STAR [12],
ALICE [13--18], and CMS Collaborations [19]. The kinetic
freeze-out temperature of interacting system and transverse flow
velocity of final-state particles are then extracted from the
comparisons and analyses.
\\

{\section{The model and formulism}}

In this paper, the transverse momentum spectra are described in
the framework of the multisource thermal model [8--10], which
assumes that many emission sources are formed in high energy
collisions and are separated into a few groups resulting from
different interacting mechanisms in the collisions as well as
different event samples in experiment measurements. The sources in
the same group have the same excitation degree and stay at a
common local equilibrium state, which can be described by using
different distribution laws. The final-state distribution is
attributed to all sources in different groups, which results in a
multi-temperature emission process if we use the standard
distribution. This also means that the transverse momentum spectra
can be described by a multi-component standard distribution which
can be fitted by the Tsallis distribution. In fact, we can adopt
one-component Tsallis-standard (T-S) distribution (Tsallis form of
the Boltzmann, Fermi-Dirac, or Bose-Einstein distribution),
one-component Tsallis distribution, and two- or three-component
standard distribution (Boltzmann, Fermi-Dirac, or Bose-Einstein
distribution), to describe the transverse momentum spectra of
final-state light flavour particles.

According to refs. [3, 20], we use the uniform expressions of the
Tsallis-standard distribution and Tsallis distribution which have
more than one version [20--26], respectively. Based on the
invariant particle momentum distribution, as well as the
unit-density function of transverse momentum ($p_T$) and rapidity
($y$), the Tsallis-standard transverse momentum distribution and
the Tsallis transverse momentum distribution are derived [3, 20].
In the present work, the formalism of the Tsallis-standard
transverse momentum distribution is adopted to be [3, 20]
\begin{equation}
f_{\rm T\textrm{-}S}(p_T)=\frac{1}{N}\frac{dN}{dp_T} =C_{\rm
T\textrm{-}S0}p_T\sqrt{p_T^2+m_0^2}\int_{y_{\min}}^{y_{\max}}
\cosh y\bigg\{\bigg[1\pm\frac{q_{\rm T\textrm{-}S}-1}{T_{\rm
T\textrm{-}S}}\bigg(\sqrt{p_T^2+m_0^2} \cosh y-\mu\bigg)
\bigg]^{\pm\frac{1}{q_{\rm T\textrm{-}S}-1}}+S\bigg\}^{-1}dy,
\end{equation}
where $C_{\rm T\textrm{-}S0}$ is the normalization constant
$gV/(2\pi)^2$ which results from $\int_0^{\infty} f_{\rm
T\textrm{-}S}(p_T) dp_T=1$; $g$, $V$, $N$, $\mu$, and $m_0$ are
degeneracy factor, volume, particle number, chemical potential,
and particle's rest mass, respectively; $y_{\min}$ is the minimum
rapidity and $y_{\max}$ is the maximum rapidity; The $+$ and $-$
in the $\pm$ sign are for $\sqrt{p_T^2+m_0^2}\cosh y>\mu$ and
$\sqrt{p_T^2+m_0^2}\cosh y\leq\mu$, respectively; $S$ has values
0, $+1$, and $-1$, which denote the Boltzmann, Fermi-Dirac, and
Bose-Einstein distributions, respectively; $T_{\rm T\textrm{-}S}$
is the effective temperature of emission sources and $q_{\rm
T\textrm{-}S}$ is the entropy index or nonequilibrium degree
factor.

In the calculation, since the effect of chemical potential can be
ignored (i.e. $\mu=0$) in collisions at RHIC and LHC energies, we
have $\sqrt{p_T^2+m_0^2}\cosh y>\mu$ and take $+$ in the $\pm$
sign. The final Tsallis-standard $p_T$ distribution is given by
\begin{equation}
f_{\rm T\textrm{-}S}(p_T)=\frac{1}{N}\frac{dN}{dp_T} =C_{\rm
T\textrm{-}S0}p_T\sqrt{p_T^2+m_0^2}\int_{y_{\min}}^{y_{\max}}
\cosh y\bigg\{\bigg[1+\frac{q_{\rm T\textrm{-}S}-1}{T_{\rm
T\textrm{-}S}}\sqrt{p_T^2+m_0^2} \cosh y \bigg]^{\frac{1}{q_{\rm
T\textrm{-}S}-1}}+S\bigg\}^{-1} dy.
\end{equation}
Further, considering $S=0$ and the power index $q_{\rm
T\textrm{-}S}/(q_{\rm T\textrm{-}S}-1)$ and other limitations, the
Tsallis transverse momentum distribution is obtained and can be
written as [20]
\begin{equation}
f_{\rm T}(p_T)=\frac{1}{N}\frac{dN}{dp_T}=C_{\rm T0}
p_T\sqrt{p_T^2+m_0^2}\int_{y_{\min}}^{y_{\max}} \cosh
y\bigg[1+\frac{q_{\rm T}-1}{T_{\rm T}}\sqrt{p_T^2+m_0^2} \cosh
y\bigg]^{-\frac{q_{\rm T}}{q_{\rm T}-1}} dy,
\end{equation}
where $C_{\rm T0}$ is the normalization constant which gives
$\int_0^{\infty} f_{\rm T}(p_T) dp_T=1$, $T_{\rm T}$ is the
effective temperature, and $q_{\rm T}$ is the entropy index or
nonequilibrium degree factor. Sometimes the upper index $q_{\rm
T}/(q_{\rm T}-1)$ is replaced by $1/(q_{\rm T}-1)$ due to $q_{\rm
T}$ being very close to 1 and application of mean field
approximation. The latter obtains a smaller $q_{\rm T}$.

Particularly, in the present work, we also use the two- or
three-component standard distribution, which is different from the
distributions introduced above. The standard Boltzmann,
Fermi-Dirac, and Bose-Einstein distributions for the $i$th
component (group) can be uniformly shown as
\begin{equation}
f_{i}(p_T)=\frac{1}{N}\frac{dN}{dp_T}=C_{i0}
p_T\sqrt{p_T^2+m_0^2}\int_{y_{\min}}^{y_{\max}} \cosh
y\bigg[\exp\bigg(\frac{\sqrt{p_T^2+m_0^2} \cosh
y}{T_{i}}\bigg)+S\bigg]^{-1} dy,
\end{equation}
where $C_{i0}$ is the normalization constant which gives
$\int_0^{\infty} f_i(p_T) dp_T=1$ and $T_{i}$ is the effective
temperature for the $i$th component. In final state, the $p_{T}$
spectrum is contributed by the $l$ components of the distribution;
that is
\begin{equation}
f_{\rm S}(p_{T})= \frac{1}{N}\frac{dN}{dp_{T}}=
\sum_{i=1}^{l}k_{i}f_{i}(p_{T}),
\end{equation}
where $k_{i}$ is the relative weight contributed by the $i$th
component. This is the multi-component standard distribution.
Considering the relative contribution of each component, we have
the mean effective temperature to be
\begin{equation}
T_{\rm S}=\sum_{i=1}^{l}k_{i}T_{i},
\end{equation}
where $T_{\rm S}$ reflects the mean excitation degree for
different components and can be used to describe the effective
temperature of emission source. We would like to point out that
Eq. (6) is not a simple additive treatment for different
temperatures, but an average weighted by different $k_i$, where
$\sum_{i=1}^{l}k_{i}=1$.

It is expected that we can obtain the relation between the
effective temperature $T$ ($T_{\rm T\textrm{-}S}$, $T_{\rm T}$, or
$T_{\rm S}$) and the particle's rest mass $m_0$. A linear fitting
can obtain the intercept $T_0$ ($T_{\rm T\textrm{-}S0}$, $T_{\rm
T0}$, or $T_{\rm S0}$) in linear relation $T-m_0$. As the
temperature corresponds to massless ($m_0=0$) particle, $T_0$ is
regarded as the source real temperature, or the kinetic freeze-out
temperature of interacting system. According to refs. [11,
27--30], we have the relation between $T$ and $T_0$,
\begin{equation}
T=T_{0}+am_{0}.
\end{equation}
According to refs. [27, 28], the slope $a$ can be given by
$v_0^2/2$ and $v_{0}$ is the (average and transverse) radial flow
velocity which is valid for low $p_T$ only. Considering different
distribution laws, $v_0$ can be $v_{\rm T\textrm{-}S0}$, $v_{\rm
T0}$, or $v_{\rm S0}$, and $T_0$ can be $T_{\rm T\textrm{-}S0}$,
$T_{\rm T0}$, or $T_{\rm S0}$, corresponding to Tsallis-standard,
Tsallis, or standard distribution, respectively. In the above
discussions, different distribution laws are in fact regarded as
different types of ``thermometers" or ``thermometric scale" and
``speedometers".

Since $a=v_0^2/2$ in Eq. (7) is only valid for low $p_T$ region
[27, 28], whereas the effective temperature is extracted in the
present work for fits in $p_T$ range which goes beyond 2 GeV/$c$,
we give up to extract radial flow $v_0$ from the slope $a$ in Eq.
(7). Although other works [11, 29, 30] also regard the intercept
$T_0$ in Eq. (7) as the kinetic freeze-out temperature, different
relations between radial flow velocity $v_0$ and slope $a$ are
used, and in some cases the relation is undetermined. In view of
uncertain relations between radial flow velocity and slope in wide
$p_T$ range, we give up to extract radial flow velocity from the
slope, and extract only the kinetic freeze-out temperature from
$T_0$ in Eq. (7).

We need an alternative method to extract mean (transverse) flow
velocity. In our very recent work [31], the linear relations
between $T$ and $m_0$ [Eq. (7)], mean transverse momentum $\langle
p_T \rangle$ and $m_0$, mean momentum $\langle p \rangle$ and
$m_0$, $T$ and mean moving mass $\overline{m}$, $\langle p_T
\rangle$ and $\overline{m}$, as well as $\langle p \rangle$ and
$\overline{m}$ are studied. From the analyses on dimension and
quantity, we regard the intercept in $T-m_0$ relation [Eq. (7)] as
the kinetic freeze-out temperature and the slope in $\langle p_T
\rangle - \overline{m}$ (or $\langle p \rangle - \overline{m}$)
relation as the mean transverse flow velocity $\langle u_T
\rangle$ (or mean flow velocity $\langle u \rangle$).
Particularly, for $\langle p_T \rangle - \overline{m}$ and
$\langle p \rangle - \overline{m}$ relations, we have
\begin{equation}
\langle p_T \rangle = \langle p_T \rangle_0 + \langle u_T \rangle
\overline{m}
\end{equation}
and
\begin{equation}
\langle p \rangle = \langle p \rangle_0 + \langle u \rangle
\overline{m},
\end{equation}
where $\langle p_T \rangle_0$ and $\langle p \rangle_0$ denote the
mean transverse momentum and mean momentum of massless particle,
respectively. We would like to point out that, in the case of
describing meanwhile the same experimental data, the three
distributions result in nearly the same $\langle p_T \rangle$ (or
$\langle p \rangle$), the same $\langle p_T \rangle -
\overline{m}$ (or $\langle p \rangle - \overline{m}$) relation,
and the same $\langle u_T \rangle$ (or $\langle u \rangle$).

From the above analyses, we see that the linear relations between
$T$ and $m_0$, $\langle p_T \rangle$ and $\overline{m}$, as well
as $\langle p \rangle$ and $\overline{m}$ [Eqs. (7)--(9)] can be
respectively obtained from the same set of parameter values which
are extracted from the same set of experimental data. This means
that the extraction processes of $T_0$, $\langle u_T \rangle$, and
$\langle u \rangle$ are independent, though their values are
entangled due to the same set of parameters. In particular, $v_0$
and the slope in Eq. (7) are related to (transverse) flow
velocity. However, there is no obvious and exact relation between
them [11, 27--30]. Anyhow, we think that the intercept in Eq. (7),
the slope in Eq. (8), and the slope in Eq. (9) can provide a set
of alternative methods to extract the kinetic freeze-out
temperature, transverse flow velocity, and flow velocity,
respectively [31]. Thus, we can use this set of alternative
methods in the present work to extract separately the kinetic
freeze-out temperature and transverse flow velocity according to
transverse momentum spectra. This set of alternative methods is
also used to judge different kinetic freeze-out scenarios in our
another recent work [32] in which an evidence of mass-dependent
differential kinetic freeze-out scenario is observed, while the
single and double kinetic freeze-out scenarios are eliminated.
\\

\begin{figure}
\hskip-1.0cm \begin{center}
\includegraphics[width=12.0cm]{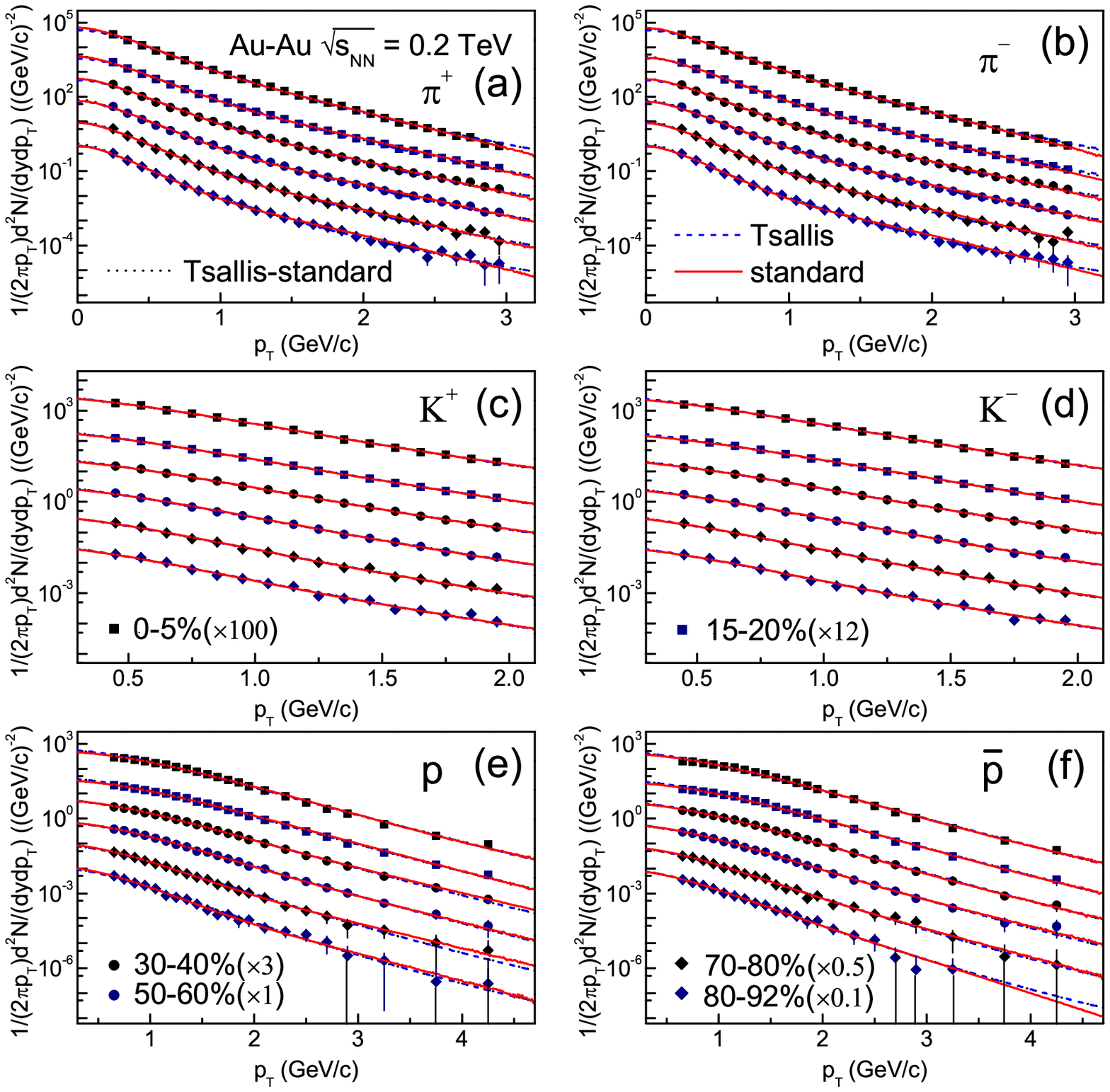}
\end{center}
\vskip-1.5cm Figure 1. Transverse momentum spectra for (a)
$\pi^+$, (b) $\pi^-$, (c) $K^+$, (d) $K^-$, (e) $p$, and (f)
$\bar{p}$ produced in Au-Au collisions with different centrality
bins at $\sqrt{s_{NN}}=0.2$ TeV. The data are measured by the
PHENIX Collaboration at midrapidity [11] and are scaled vertically
as quoted in the figure. The dotted, dashed, and solid curves are
our results calculated by using the Tsallis-standard, Tsallis, and
two-component standard distributions, respectively.
\end{figure}

{\section{Results and discussion}}

Figure 1 presents the centrality dependence of $p_{T}$ spectra for
(a) $\pi^+$, (b) $\pi^-$, (c) $K^+$, (d) $K^-$, (e) $p$, and (f)
$\bar{p}$ produced in Au-Au collisions at center-of-mass energy
$\sqrt{s_{NN}}=0.2$ TeV. The data measured by the PHENIX
Collaboration at midrapidity [11] are represented by different
symbols which correspond to different centrality ($C$) bins from
0--5\% to 80--92\% scaled by different amounts as shown in the
panels for clarity. The error bars are statistical only. The
dotted, dashed, and solid curves are our results calculated by
using the Tsallis-standard, Tsallis, and two-component standard
distributions, respectively. The values of free parameters
($T_{\rm T\textrm{-}S}$, $q_{\rm T\textrm{-}S}$, $T_{\rm T}$, and
$q_{\rm T}$), normalization constants ($N_{\rm T\textrm{-}S0}$ and
$N_{\rm T0}$) for comparisons between curves and data, and
$\chi^2$ per degree of freedom ($\chi^2$/dof) for Tsallis-standard
and Tsallis distributions are given in Table 1. The values of free
parameters ($T_{1}$, $k_{1}$, and $T_{2}$), mean effective
temperature ($T_{\rm S}$), normalization constant ($N_{\rm S0}$)
for comparisons between curves and data, and $\chi^2$/dof for
two-component standard distribution are listed in Table 2. One can
see that all three types of distribution laws are consistent with
the experimental data. The effective temperatures $T_{\rm
T\textrm{-}S}$, $T_{\rm T}$, and $T_{\rm S}$ increase with the
increase of centrality or particle mass, and $T_{\rm T\textrm{-}S}
\leq T_{\rm T}<T_{\rm S}$ for a given set of data. We would like
to point out that the increase of centrality and the decrease of
centrality percentage have the same meaning.

\begin{figure}
\hskip-1.0cm \begin{center}
\includegraphics[width=12.0cm]{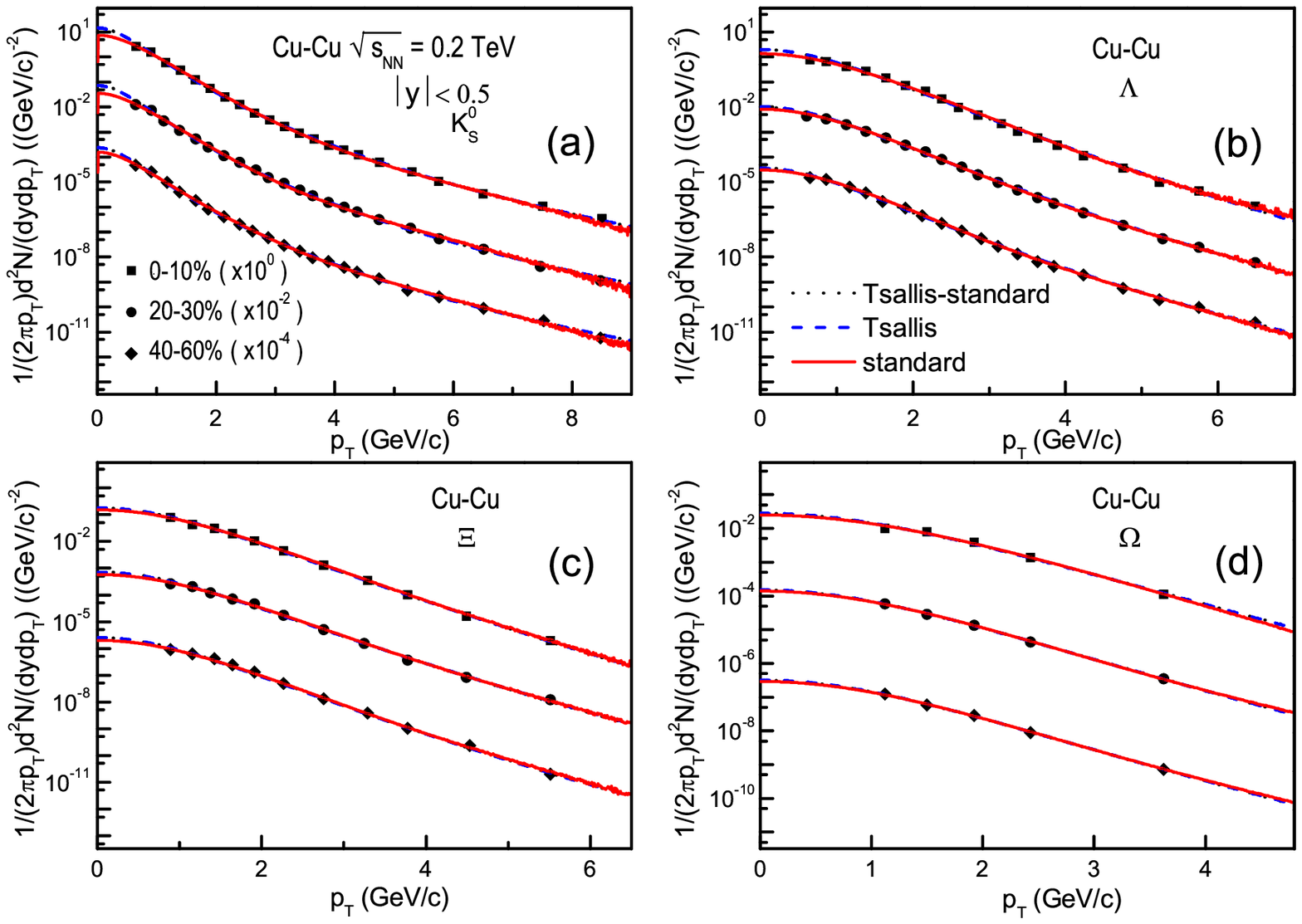}
\end{center}
\vskip-1.0cm Figure 2. Transverse momentum spectra of (a)
$K_{S}^{0}$, (b) $\Lambda$, (c) $\Xi$, and (d) $\Omega$ produced
in Cu-Cu collisions at $\sqrt{s_{NN}}=0.2$ TeV in three centrality
intervals. The symbols represent the experimental data recorded by
the STAR Collaboration in the rapidiy range $|y|<0.5$ [12]. The
dotted, dashed, and solid curves are our results calculated by
using the Tsallis-standard, Tsallis, and two- or three-component
standard distributions, respectively.
\end{figure}

Figure 2 shows $p_{T}$ spectra of (a) $K_{S}^{0}$, (b) $\Lambda$,
(c) $\Xi$, and (d) $\Omega$ produced in Cu-Cu collisions at
$\sqrt{s_{NN}}=0.2$ TeV in different centrality intervals of
0--10\%, 20--30\%, and 40--60\%. The symbols represent the
experimental data recorded by the STAR Collaboration in the
rapidiy range $|y|<0.5$ [12]. For clarity, the results for
different $C$ intervals are scaled by different amounts shown in
the panels. The uncertainties on the data points include both
statistical and systematic errors. The dotted, dashed, and solid
curves are our results based on the Tsallis-standard, Tsallis, and
two- or three-component standard transverse momentum
distributions, respectively. The values of free parameters
($T_{\rm T\textrm{-}S}$, $q_{\rm T\textrm{-}S}$, $T_{\rm T}$,
$q_{\rm T}$, $T_1$, $k_1$, $T_2$, $k_2$, and $T_3$), $T_{\rm S}$,
normalization constants ($N_{\rm T\textrm{-}S0}$, $N_{\rm T0}$,
and $N_{\rm S0}$), and $\chi^2$/dof are displayed in Tables 1 and
2. Obviously, the experimental data can be described by the three
types of fit functions for $p_{T}$ in all centrality bins. The
effective temperatures $T_{\rm T\textrm{-}S}$, $T_{\rm T}$, and
$T_{\rm S}$ increase with the increase of centrality or particle
mass, and $T_{\rm T\textrm{-}S} \leq T_{\rm T}<T_{\rm S}$ for a
given set of data.

\begin{figure}
\hskip-1.0cm \begin{center}
\includegraphics[width=12.0cm]{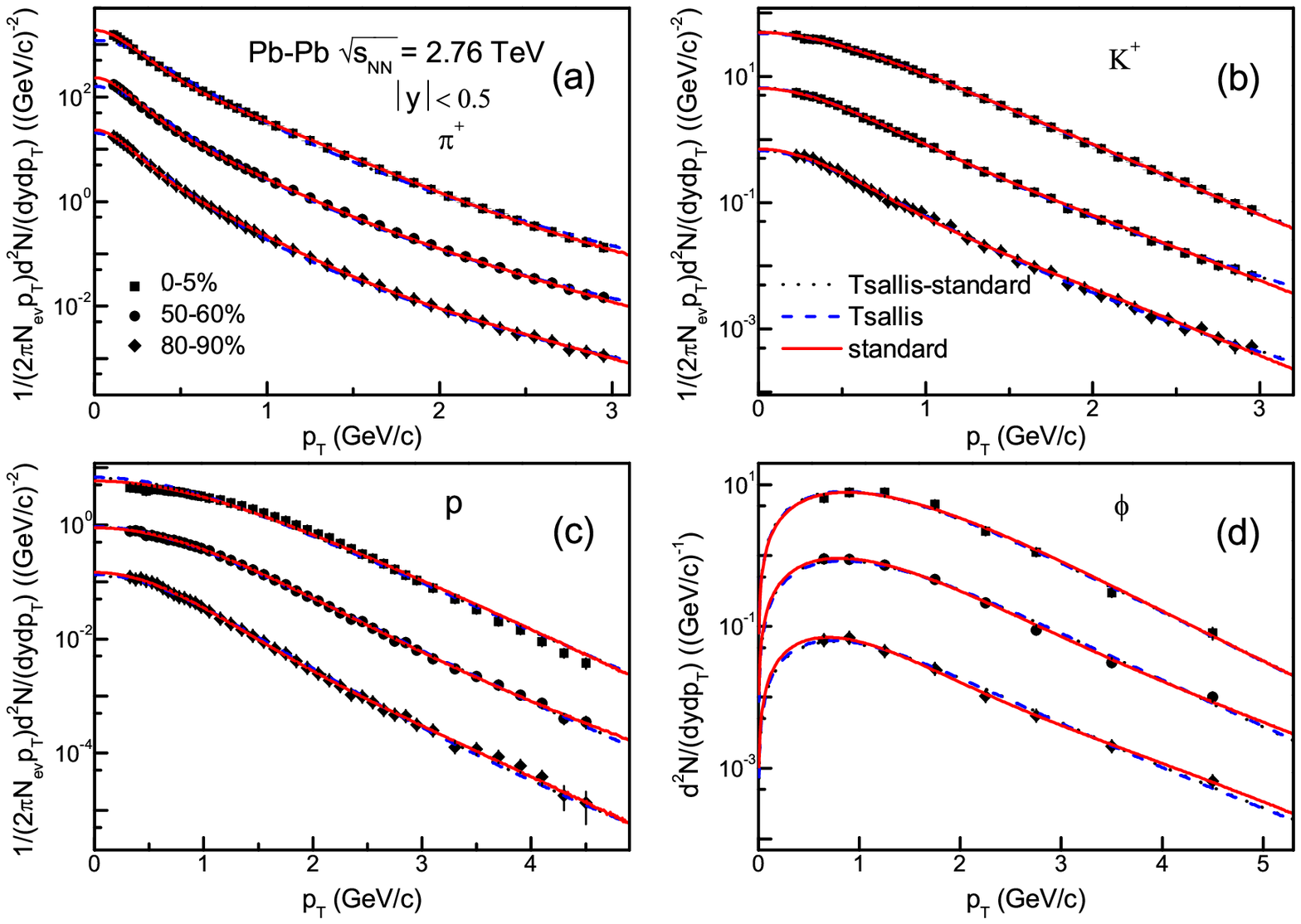}
\end{center}
\vskip-1.0cm Figure 3. Transverse momentum spectra of (a)
$\pi^{+}$, (b) $K^{+}$, (c) $p$, and (d) $\phi$ produced in Pb-Pb
collisions at $\sqrt{s_{NN}}=2.76$ TeV in three centrality
intervals. The symbols represent the experimental data measured by
the ALICE Collaboration at mid-rapidity ($|y|<0.5$) [13, 14]. The
fitting results with three types of distributions are plotted by
the curves.
\end{figure}

\begin{figure}
\hskip-1.0cm \begin{center}
\includegraphics[width=12.0cm]{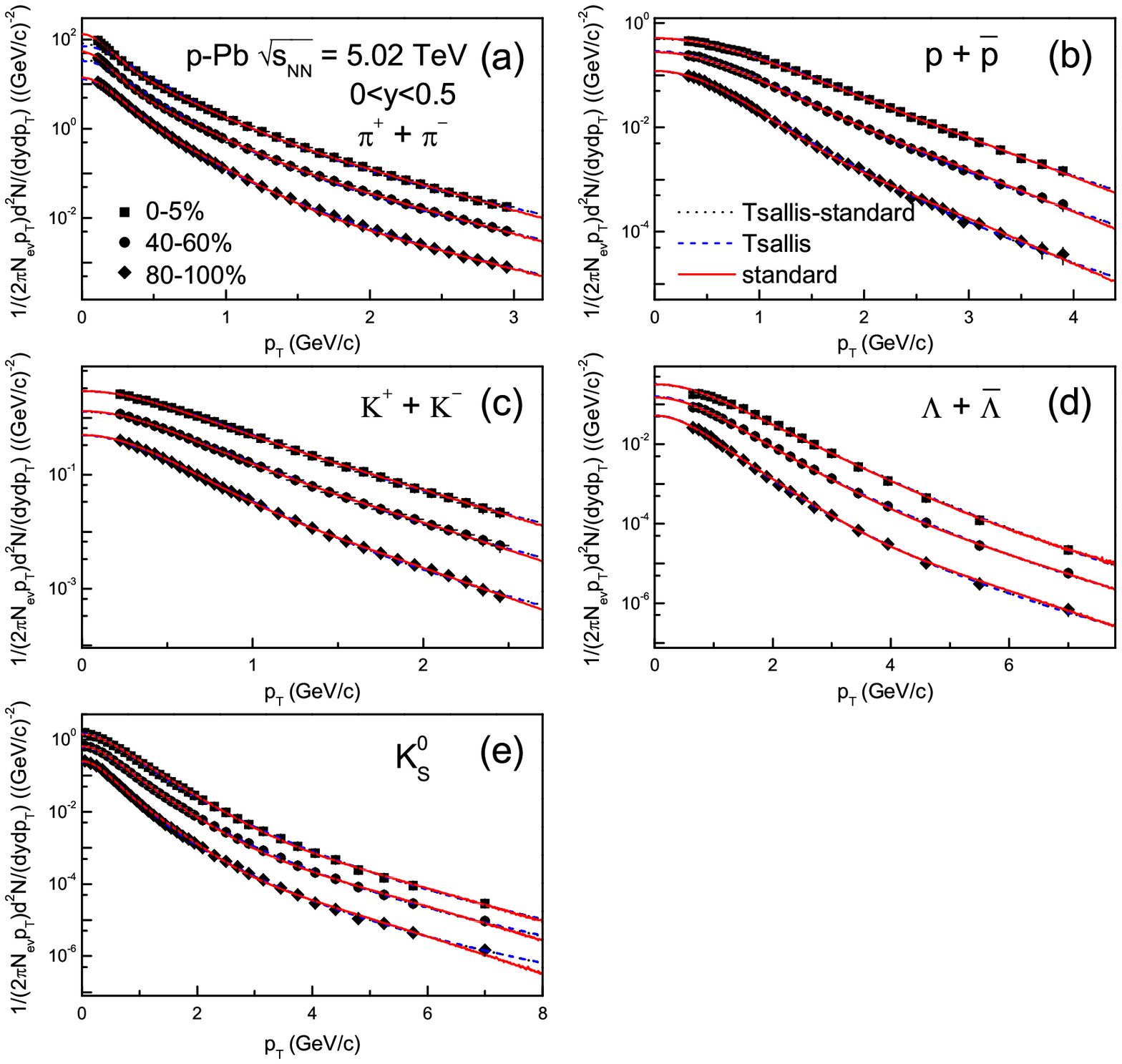}
\end{center}
\vskip-1.5cm Figure 4. Transverse momentum spectra of (a)
$\pi^{\pm}$, (b) $p+\bar{p}$, (c) $K^{\pm}$, (d) $\Lambda +
\bar{\Lambda}$, and (e) $K_{S}^{0}$ in $p$-Pb collisions in
different centrality classes at $\sqrt{s_{NN}}=5.02$ TeV. The
experimental data represented by the symbols are performed by the
ALICE Collaboration in $0<y<0.5$ [15]. Our fitting results are
exhibited by the curves.
\end{figure}

The $p_{T}$ spectra of (a) $\pi^{+}$, (b) $K^{+}$, (c) $p$, and
(d) $\phi$ produced in central (0--5\%), semi-central (50--60\%),
and peripheral (80--90\%) Pb-Pb collisions at $\sqrt{s_{NN}}=2.76$
TeV are displayed in Fig. 3, where $N_{\rm ev}$ denotes the number
of events. The data points are measured by the ALICE Collaboration
at midrapidity ($|y|<0.5$) [13, 14]. The error bars combined both
statistical and systematic uncertainties. The fitting results with
Tsallis-standard, Tsallis, and two- or three-component standard
distributions are plotted by the dotted, dashed, and solid curves,
respectively. The values of free parameters, $T_{\rm S}$,
normalization constants, and $\chi^2$/dof are shown in Tables 1
and 2. These fitting functions describe the experimental data.
Once again, the effective temperatures $T_{\rm T\textrm{-}S}$,
$T_{\rm T}$, and $T_{\rm S}$ increase with increase of centrality
or particle mass, and $T_{\rm T\textrm{-}S} \leq T_{\rm T}<T_{\rm
S}$ for a given set of data.

Figure 4 exhibits $p_{T}$ spectra of (a) $\pi^{\pm}$, (b) $p +
\bar{p}$, (c) $K^{\pm}$, (d) $\Lambda +\bar{\Lambda}$, and (e)
$K_{S}^{0}$ produced in $p$-Pb collisions in different centrality
classes (0--5\%, 40--60\%, and 80--100\%) at $\sqrt{s_{NN}}=5.02$
TeV. The experimental data represented by different symbols are
performed by the ALICE Collaboration in the rapidity interval
$0<y<0.5$ [15]. The statistical and systematic uncertainties are
combined in the error bars. Our results from the analyses of the
Tsallis-standard, Tsallis, and two- or three-component standard
distributions are exhibited by the dotted, dashed, and solid
curves, respectively. The values of free parameters, $T_{S}$,
normalization constants, and $\chi^2$/dof are summarized in Tables
1 and 2. As can be seen, the fitting results are in agreement with
the experimental data. The effective temperatures $T_{\rm
T\textrm{-}S}$, $T_{\rm T}$, and $T_{\rm S}$ increase with the
increase of centrality or particle mass, and $T_{\rm T\textrm{-}S}
\leq T_{\rm T}<T_{\rm S}$ for a given set of data.

\begin{figure}
\hskip-1.0cm \begin{center}
\includegraphics[width=12.0cm]{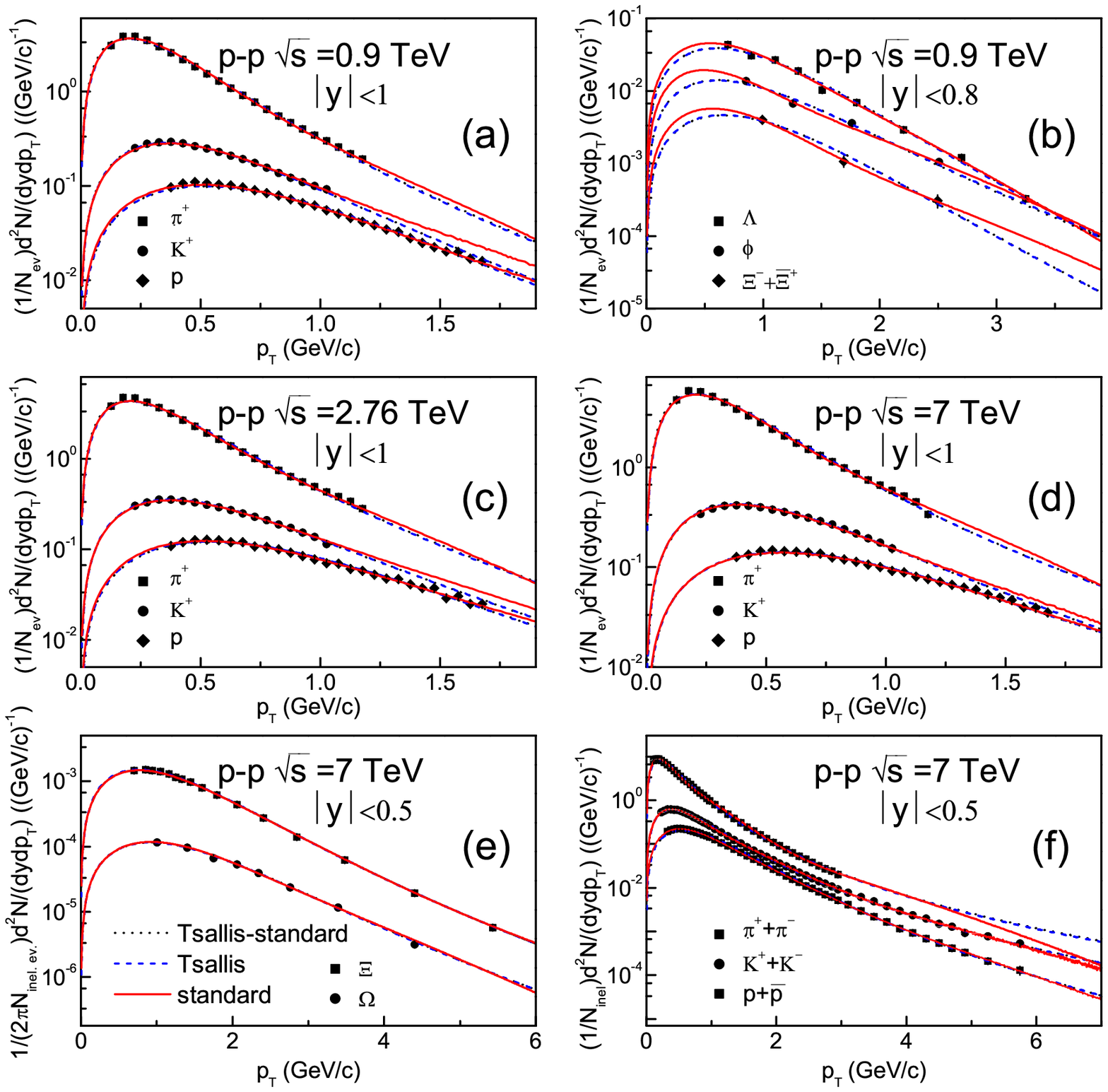}
\end{center}
\vskip-1.5cm Figure 5. Transverse momentum spectra of various
identified hadrons produced in $p$-$p$ collisions. The symbols in
Figs. 5(a), 5(c), and 5(d) represent the experimental data for
$\pi^{+}$, $K^{+}$, and $p$ measured by the CMS Collaboration [19]
in the range $|y|<1$ at $\sqrt{s}=0.9, 2.76, 7$ TeV, respectively.
The data points of the ALICE Collaboration measured in $|y|<0.8$
at $\sqrt{s}=0.9$ TeV for $\Lambda$, $\phi$, and
$\Xi^{-}+\bar{\Xi}^{+}$ [16], in $|y|<0.5$ at $\sqrt{s}=7$ TeV for
$\Xi$ and $\Omega$ [17], and in $|y|<0.5$ at $\sqrt{s}=7$ TeV for
$\pi^{\pm}$, $K^{\pm}$, and $p$+$\bar{p}$ [18] are shown in Figs.
5(b), 5(e), and 5(f), respectively. The fitting curves are our
results.
\end{figure}

Figure 5 gives $p_{T}$ spectra of various identified hadrons
produced in $p$-$p$ collisions at different energies, where
$N_{\rm inel}$ denotes the number of inelastic events. The symbols
in Figs. 5(a), 5(c), and 5(d) represent the experimental data for
$\pi^{+}$, $K^{+}$, and $p$ measured by the CMS Collaboration [19]
in the range $|y|<1$ at $\sqrt{s}=0.9$, 2.76, and 7 TeV,
respectively. The data points of the ALICE Collaboration measured
in $|y|<0.8$ at $\sqrt{s}=0.9$ TeV for $\Lambda$, $\phi$, and
$\Xi^{-}+\bar{\Xi}^{+}$ [16], in $|y|<0.5$ at $\sqrt{s}=7$ TeV for
$\Xi$ and $\Omega$ [17], and in $|y|<0.5$ at $\sqrt{s}=7$ TeV for
$\pi^{\pm}$, $K^{\pm}$, and $p$+$\bar{p}$ [18] are shown in Figs.
5(b), 5(e), and 5(f), respectively. The uncertainties
corresponding to combined statistics and systematics are shown as
error bars. The fitting results (dotted, dashed, and solid curves)
using mentioned functional forms (Tsallis-standard, Tsallis, and
two- or three-component standard distributions) are superimposed.
The values of free parameters, $T_{S}$, normalization constants,
and $\chi^2$/dof are exhibited in Tables 1 and 2. Once more, the
Tsallis-standard, Tsallis, and two- or three-component standard
distributions can describe the experimental data of the considered
particles. The effective temperatures $T_{\rm T\textrm{-}S}$,
$T_{\rm T}$, and $T_{\rm S}$ increase with the increase of
particle mass, and $T_{\rm T\textrm{-}S} \leq T_{\rm T}<T_{\rm S}$
for a given set of data.

To see clearly the dependences of effective temperatures $T_{\rm
T\textrm{-}S}$, $T_{\rm T}$, and $T_{\rm S}$ on centrality and
particle mass, the related values listed in Tables 1 and 2, which
are extracted from figures 1, 2, 3, and 4, are analyzed in Figs.
6, 7, 8 (left panel), and 8 (right panel), respectively. Different
symbols correspond to different centrality intervals shown in the
panels. The lines are our fitted results. At the same time, Fig. 9
shows the dependences of effective temperatures on particle mass
for $p$-$p$ collisions at the LHC energies. Different symbols
correspond to different effective temperatures which are listed in
Tables 1 and 2 and extracted from Fig. 5. The lines are our fitted
results. Conclusions obtained from Figs. 1--5 and Tables 1 and 2
can be clearly seen from Figs. 6--9.

\begin{figure}
\hskip-1.0cm \begin{center}
\includegraphics[width=12.0cm]{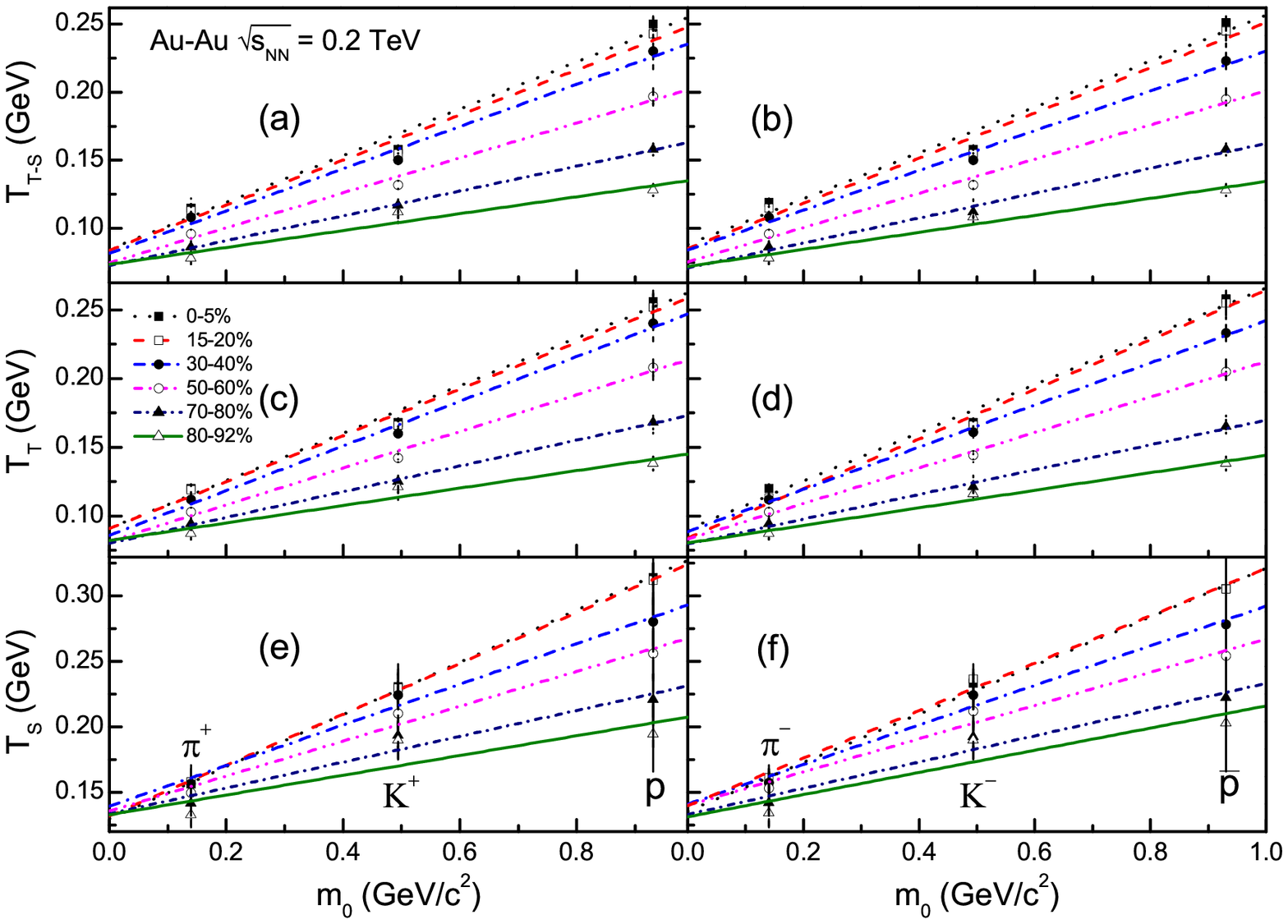}
\end{center}
\vskip-.70cm Figure 6. Particle mass and centrality dependences of
$T_{\rm T\textrm{-}S}$, $T_{\rm T}$, and $T_{\rm S}$ for positive
[left: (a), (c), and (e)] and negative particles [right: (b), (d),
and (f)] in Au-Au collisions at $\sqrt{s_{NN}}=0.2$ TeV. The
symbols represent the parameter values extracted from Fig. 1 and
listed in Tables 1 and 2. The lines represent linear fits to the
results from each centrality bin as a function of mass using Eq.
(7).
\end{figure}

\begin{figure}
\hskip-1.0cm \begin{center}
\includegraphics[width=6.0cm]{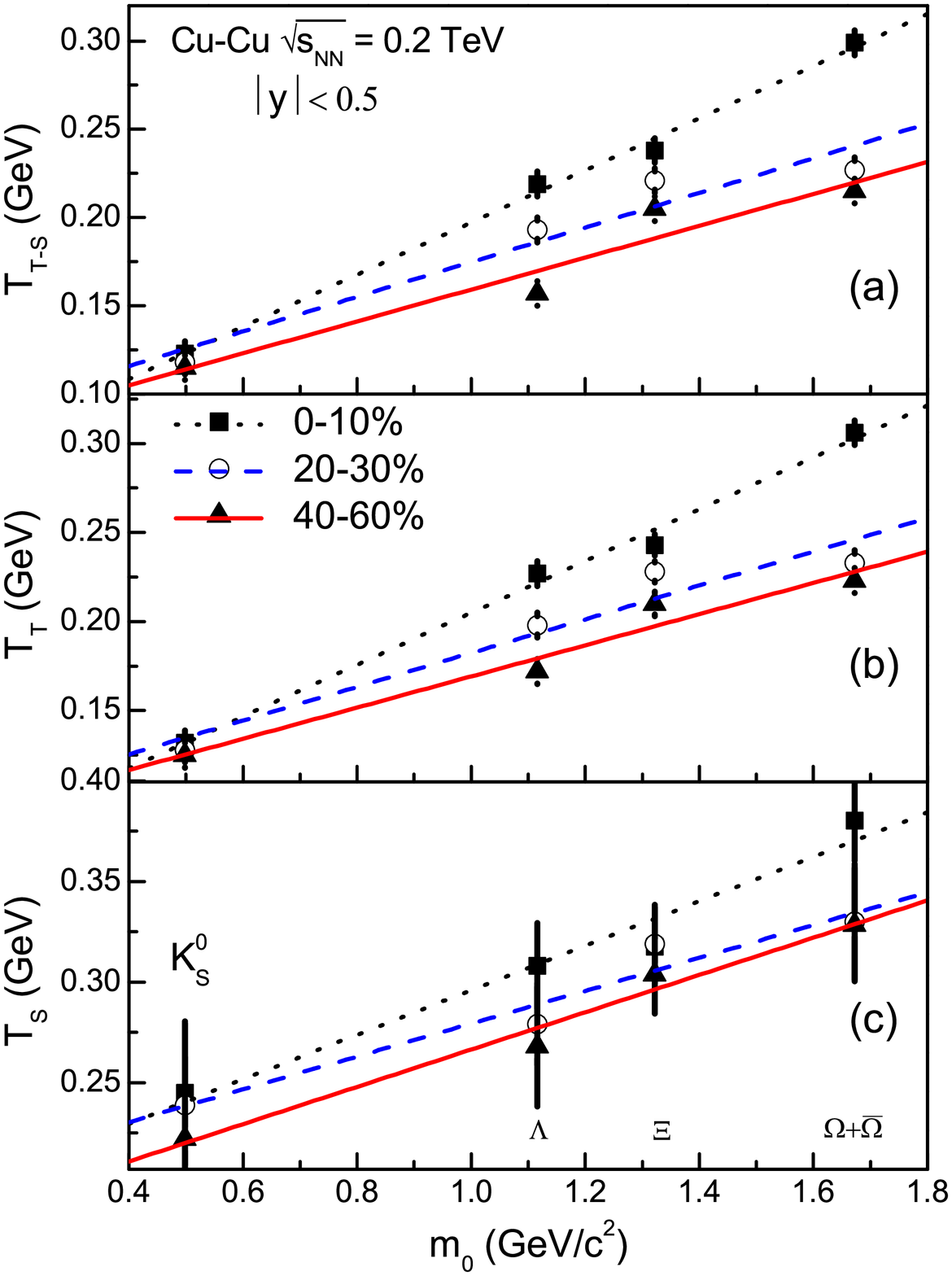}
\end{center}
\vskip-.0cm Figure 7. Same as Fig. 6, but for dependences of (a)
$T_{\rm T\textrm{-}S}$, (b) $T_{\rm T}$, and (c) $T_{\rm S}$ on
particle mass and centrality in Cu-Cu collisions at
$\sqrt{s_{NN}}=0.2$ TeV. The symbols represent the parameter
values extracted from Fig. 2.
\end{figure}

\begin{figure}
\hskip-1.0cm \begin{center}
\includegraphics[width=12.0cm]{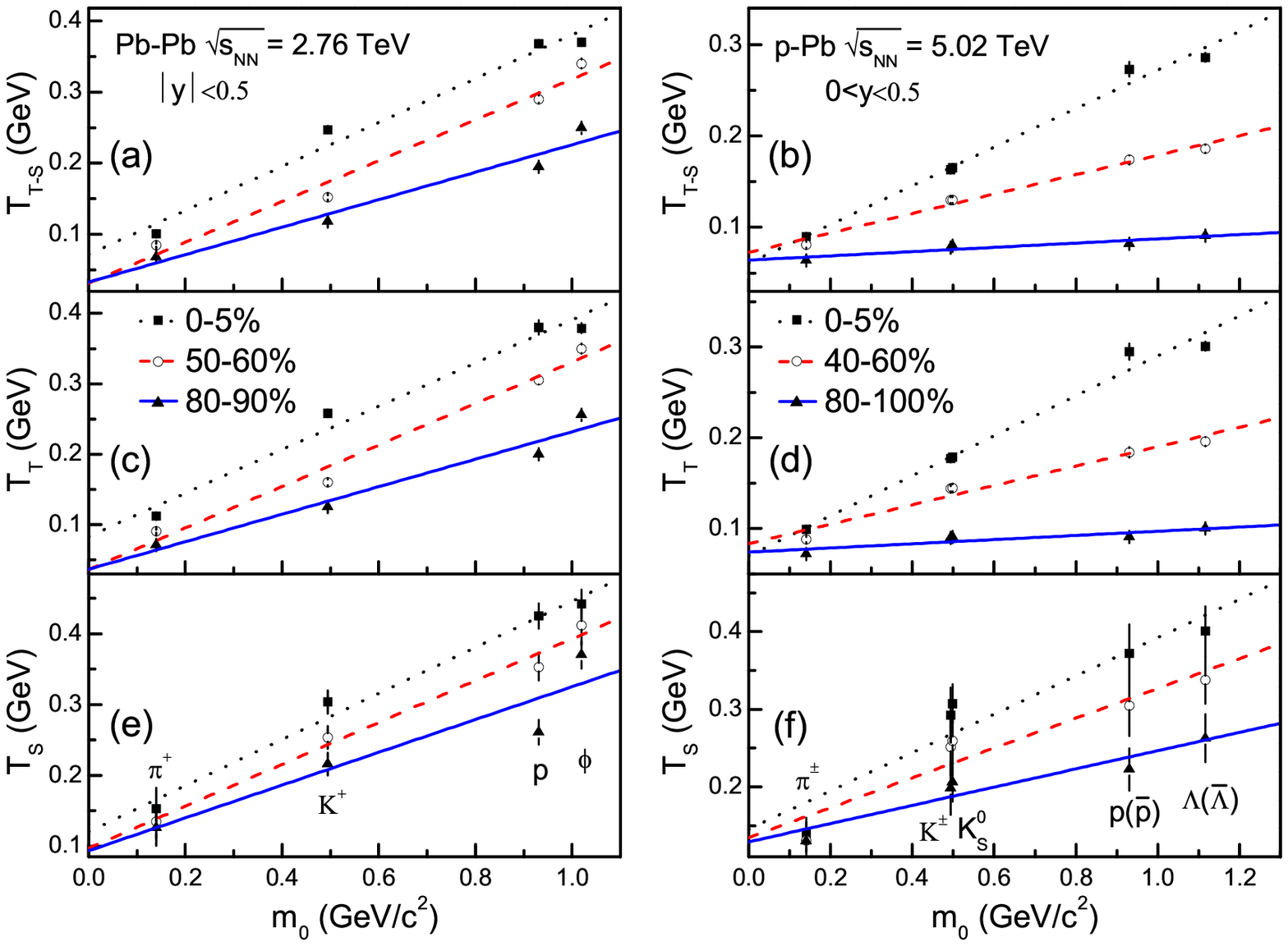}
\end{center}
\vskip-.50cm Figure 8. Same as Fig. 6, but for particle mass and
centrality dependences of $T_{\rm T\textrm{-}S}$, $T_{\rm T}$, and
$T_{\rm S}$ for particles in Pb-Pb collisions at
$\sqrt{s_{NN}}=2.76$ TeV [left: (a), (c), and (e)], and in $p$-Pb
collisions at $\sqrt{s_{NN}}=5.02$ TeV [right: (b), (d), and (f)].
The symbols represent the parameter values extracted from Figs. 3
and 4.
\end{figure}

\begin{figure}
\hskip-1.0cm \begin{center}
\includegraphics[width=12.0cm]{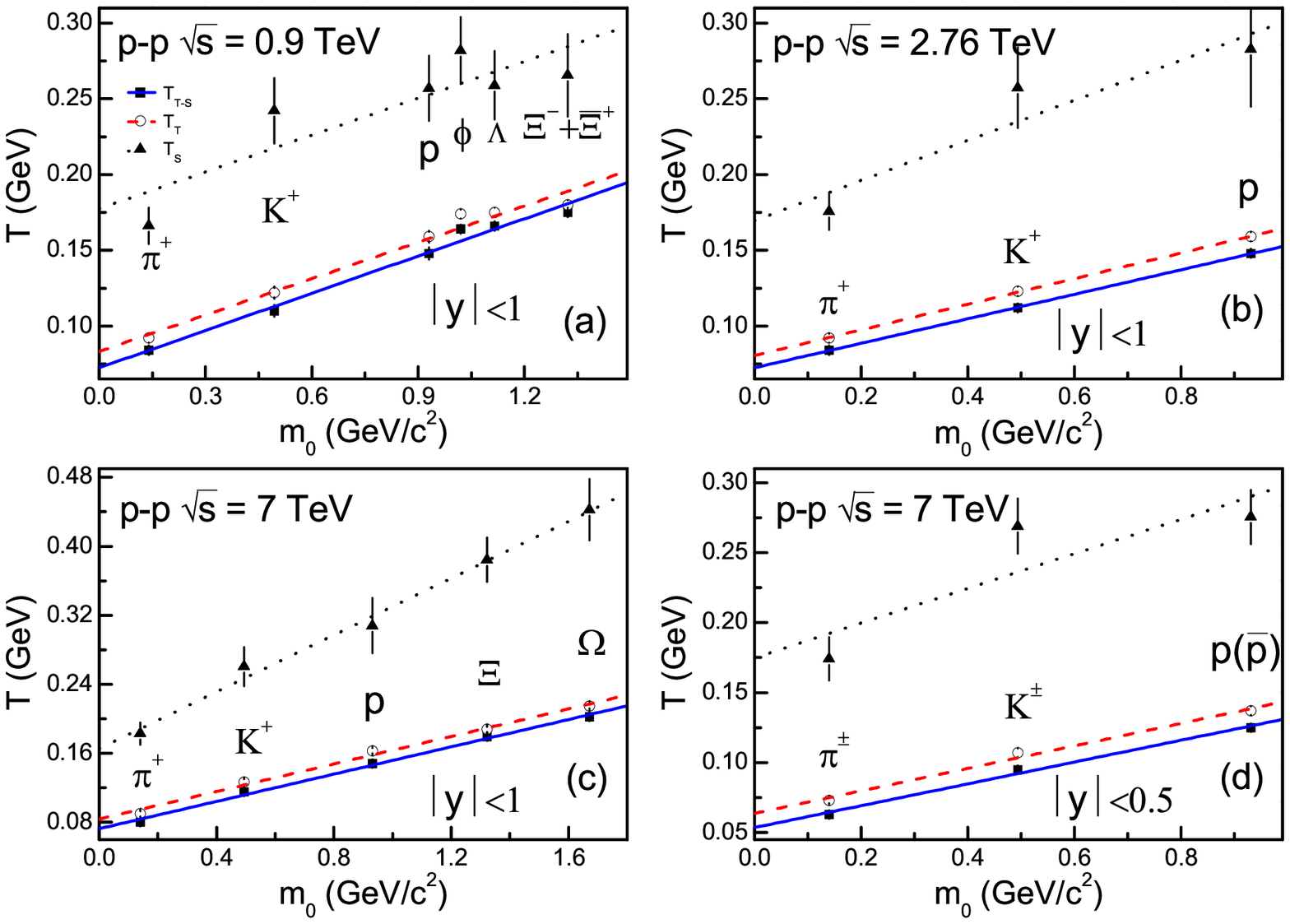}
\end{center}
\vskip-.50cm Figure 9. Same as Fig. 6, but for dependences of
$T_{\rm T\textrm{-}S}$, $T_{\rm T}$, and $T_{\rm S}$ on particle
mass in $p$-$p$ collisions at (a) $\sqrt{s}=0.9$ TeV, (b)
$\sqrt{s}=2.76$ TeV, (c) $\sqrt{s}=7$ TeV for positive particles,
and (d) $\sqrt{s}=7$ TeV for charged particles. The symbols
represent the parameter values extracted from Fig. 5.
\end{figure}

\begin{figure}
\hskip-1.0cm \begin{center}
\includegraphics[width=12.0cm]{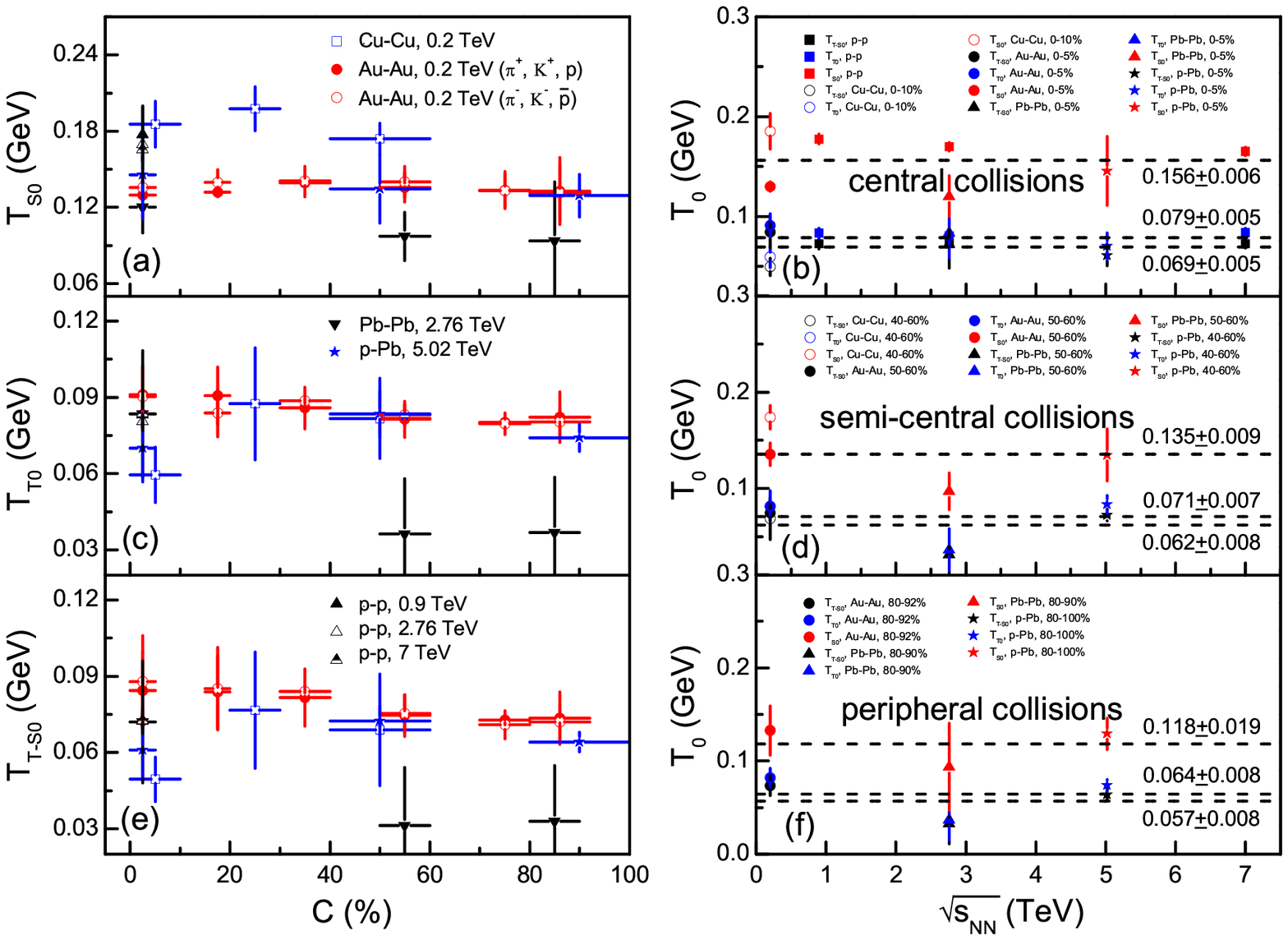}
\end{center}
\vskip-.50cm Figure 10. Centrality, energy, and size dependences
of kinetic freeze-out temperatures $T_{\rm T\textrm{-}S0}$,
$T_{\rm T0}$, and $T_{\rm S0}$. The left panel [(a), (c), and (e)]
is mainly for different kinetic freeze-out temperatures and
centralities, and the right panel [(b), (d), and (f)] is mainly
for three centralities and different energies. The symbols
represent our fitted results from Figs. 6--9. The results for
$p$-$p$ collisions are plotted for comparisons with central
(0--5\%) nucleus-nucleus collisions. The dashed lines marked with
specific values in the right panel represent the three averages
$\langle T_{\rm T\textrm{-}S0} \rangle$, $\langle T_{\rm T0}
\rangle$, and $\langle T_{\rm S0} \rangle$ for three centralities,
where $\langle T_{\rm T\textrm{-}S0} \rangle \leq \langle T_{\rm
T0} \rangle < \langle T_{\rm S0} \rangle$.
\end{figure}

In Figs. 6--9, the lines are worthwhile for further investigation.
In fact, according to Eq. (7), the intercepts at $m_0=0$ are real
temperatures (kinetic freeze-out temperatures) $T_{\rm
T\textrm{-}S0}$, $T_{\rm T0}$, and $T_{\rm S0}$, which are
extracted from Tsallis-standard, Tsallis, and two- or
three-component standard distributions, respectively. We show the
dependences of kinetic freeze-out temperatures on centrality $C$
and energy $\sqrt{s_{NN}}$ in Fig. 10 for different sizes of
collisions, where the related values for $p$-$p$ collisions are
compared with central (0--5\%) nucleus-nucleus collisions. In most
cases, $T_{\rm T\textrm{-}S0} \leq T_{\rm T0}<T_{\rm S0}$. The
kinetic freeze-out temperatures decrease slightly with the
decrease of centrality (or with increase of centrality
percentage), and do not show an obvious change with energy and
size for the central, semi-central, and peripheral collisions. In
the considered energy range from 0.2 to 7 TeV, the mean kinetic
freeze-out temperatures ($\langle ... \rangle$) extracted from
different distribution laws for the mentioned three centralities
are marked in the panels by the specific values and dashed lines.
We see that $\langle T_{\rm T\textrm{-}S0} \rangle \leq \langle
T_{\rm T0} \rangle < \langle T_{\rm S0} \rangle$, and $\langle
T_{\rm 0} \rangle_{\rm central}
> \langle T_{\rm 0} \rangle_{\rm semi\textrm{-}central} \geq
\langle T_{\rm 0} \rangle_{\rm peripheral}$. The independences of
kinetic freeze-out temperatures on energy and size render that
different interacting systems at the stage of kinetic freeze-out
reached the same excitation degree of hadronic matter in the
considered energy range.

We noticed that the value of kinetic freeze-out temperature
$T_{\rm S0}$ [($156\pm6$) MeV] in central collisions [Fig. 10(b)]
is less than that of chemical freeze-out temperature (170 MeV)
used in some theoretical estimations for the critical point of QCD
(quantum chromodynamics) [33--35]. Our result is also less than
that of kinetic freeze-out temperature (177 MeV) extracted from an
exponential shape of transverse mass spectrum [11], and equal to
that of chemical freeze-out temperature (156 MeV) extracted from
particle ratios in a thermal and statistical model [36]. The
latter item means that the kinetic freeze-out temperature obtained
in the present work is greater than that obtained from the thermal
and statistical model [36], or the kinetic and chemical
freeze-outs had happened at almost the same time. We are inclined
to the latter. These results are not exactly consistent with each
other even if we consider the kinetic freeze-out temperature being
less than the chemical freeze-out temperature. It is undoubted
that more analyses are needed in the future.

To extract the transverse flow velocity, we analyze $\langle p_T
\rangle - \overline{m}$ relations in Fig. 11, where the related
parameters listed in Tables 1 and 2 are used to calculate $\langle
p_T \rangle$ and $\overline{m}$, where the three distributions
result in the same relations. Different symbols correspond to
different centrality intervals shown in the panels. The lines are
our fitted results. From the slopes, we can obtain the mean
transverse flow velocities of particles produced in collisions
with different centralities and at different energies. The related
results are shown in Figs. 12(a) and 12(b) respectively, where the
results for $p$-$p$ collisions are shown at $C=2.5\%$ for
comparisons. One can see that the mean transverse flow velocity in
central collisions is greater than that in peripheral collisions.
The relative difference is about 10\%.

The mean transverse flow velocity $\langle u_T \rangle$
[$(0.443\pm0.071)c$] for central collisions obtained in the
present work is comparable with the radial flow velocity
[$(0.426-0.472)c$] [37, 38] of the blast-wave model [7, 37, 38],
while the value of $\langle u_T \rangle$ [$(0.403\pm0.079)c$] for
peripheral collisions obtained in the present work is far from the
radial flow velocity ($\approx 0$) [37] of the blast-wave model
[7, 37, 38]. In fact, $\langle u_T \rangle$ contains isotropic
radial flow and anisotropic collective flow (directed flow and
others). In central collisions, the effect of anisotropic
collective flow is small [39], $\langle u_T \rangle$ contains
mainly the isotropic radial flow. In peripheral collisions, the
isotropic radial flow is nearly zero [37], $\langle u_T \rangle$
contains mainly the anisotropic collective flow [39]. According to
Fig. 12(b), $\langle u_T \rangle$ in central collisions (which
contains mainly the isotropic radial flow) is greater than that in
peripheral collisions (which contains mainly the anisotropic
collective flow).

Because of $\langle u_T \rangle$ containing the isotropic radial
flow and anisotropic collective flow together, it is different
from the original blast-wave model [7, 37, 38] in which only the
isotropic radial flow is contained. Our explanation on flow
components is comparable with the improved blast-wave model [40,
41, 42] in which the third parameter is introduced to describe the
anisotropic transverse flow generated in non-central collisions
[40] and the fourth parameter is introduced to take into account
the anisotropic shape of the source in coordinate space [41].
Although the dependence of elliptic flow on transverse momentum is
described, the velocity of anisotropic transverse flow is not
available in the improved blast-wave model [40, 41, 42]. As an
indirect measurement of model independence, the present work is
hoped to cause a further discussion on the flow velocity.

\begin{figure}
\hskip-1.0cm \begin{center}
\includegraphics[width=10.0cm]{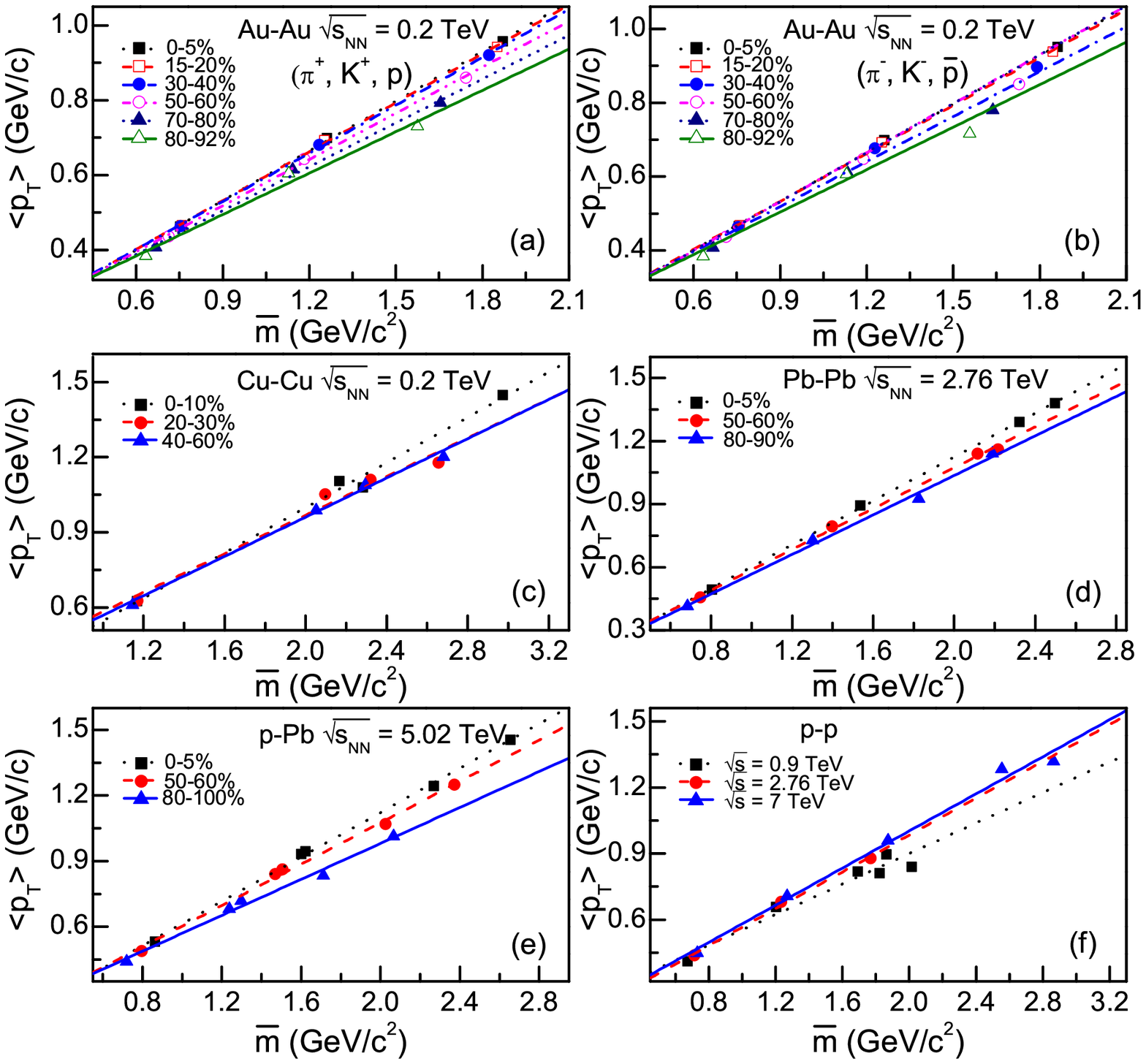}
\end{center}
\vskip-.0cm Figure 11. Dependences of $\langle p_T \rangle$ on
$\overline{m}$ in (a) 0.2 TeV Au-Au collisions for positive
particles, (b) 0.2 TeV Au-Au collisions for negative particles,
(c) 0.2 TeV Cu-Cu collisions, (d) 2.76 TeV Pb-Pb collisions, and
(e) 5.02 TeV $p$-Pb collisions with different centralities, as
well as (f) $p$-$p$ collisions at different energies. The symbols
are our calculated results due to the parameter values listed in
Tables 1 and 2. The lines are our fitted results.
\end{figure}

\begin{figure}
\hskip-1.0cm \begin{center}
\includegraphics[width=8.0cm]{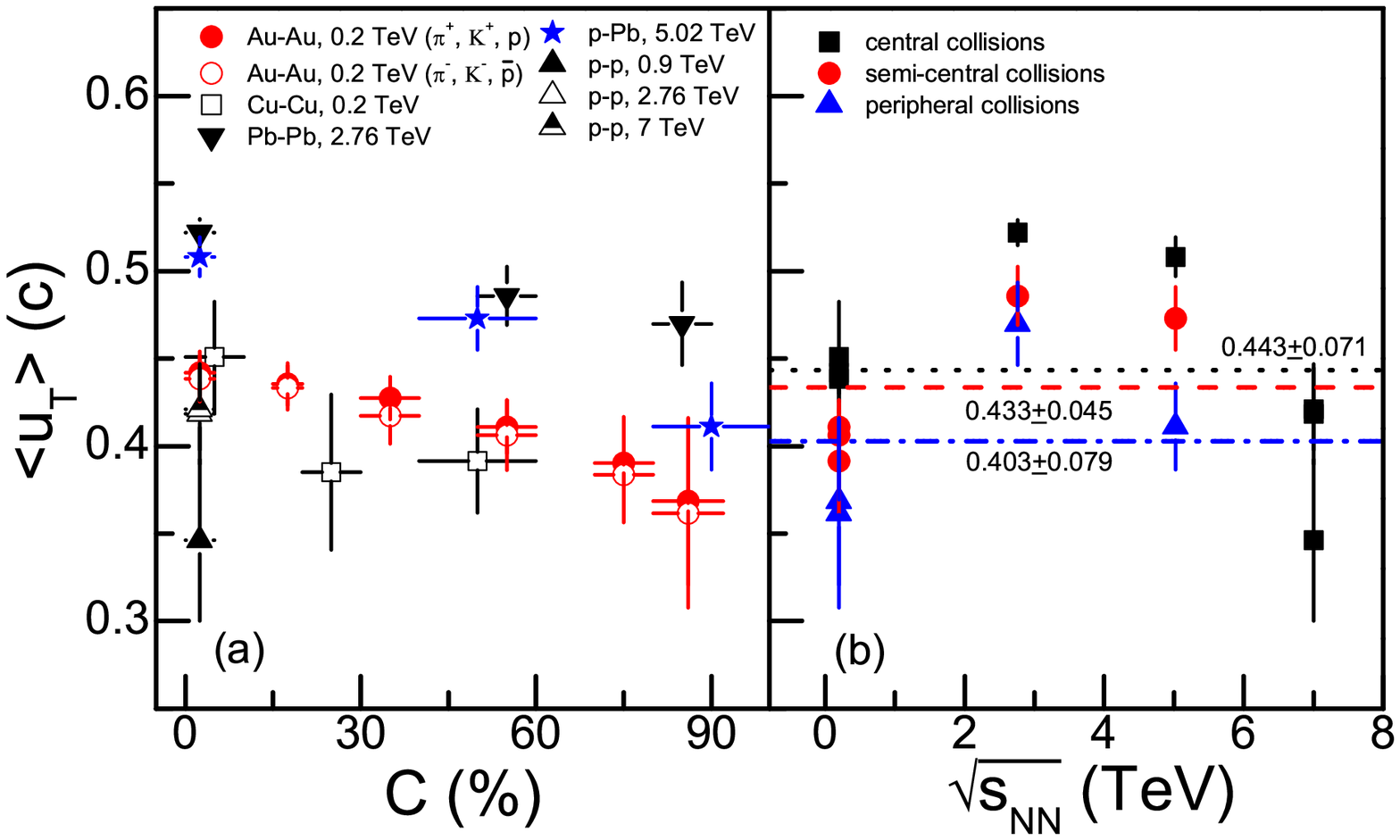}
\end{center}
\vskip-.0cm Figure 12. Dependences of $\langle u_T \rangle$ on (a)
centrality and (b) energy for different collisions marked in the
panel. The symbols are the slopes in Fig. 11. The dotted, dashed,
and dotted-dashed lines in Fig. 12(b) are the mean $\langle u_T
\rangle$ over energy for central, semi-central, and peripheral
collisions, respectively.
\end{figure}

The kinetic freeze-out temperature and transverse flow velocity
are extracted from the transverse momentum spectra, though the
methodology used in the present work is different from that of the
blast-wave model with the Boltzmann-Gibbs distribution or the
Tsallis statistics [7, 37, 38], or the quark recombination model
with the Cooper-Frye formula and flow [43]. In the latter two
models, the kinetic freeze-out temperature and radial flow
velocity can be obtained from certain formula description. In the
present work, we obtain the effective temperature from the formula
of transverse momentum distribution, and kinetic freeze-out
temperature from the intercept in linear $T-m_0$ relation [Eq.
(7)] [11, 27, 29--32] and the transverse flow velocity from the
slope in linear $\langle p_T \rangle - \overline{m}$ relation [Eq.
(8)] [32]. The method used in the present work needs two steps to
obtain the kinetic freeze-out temperature and transverse flow
velocity, though the process of calculation is simple.

Although the topic discussed by us was intensively discussed in
the literature and similar fitting and discussion were carried out
in the literature many times, we have used an alternative
methodology [11, 27, 29--32] to extract the kinetic freeze-out
temperature and transverse flow velocity, and the analysis is more
extensive than anything done before. This methodology is different
from that of the original blast-wave model [7, 37, 38] or the
quark recombination model [43]. In addition, differing from the
original blast-wave model which studies the kinetic freeze-out
temperature and isotropic radial flow, the present work focuses on
the kinetic freeze-out temperature and transverse flow which
contains together the isotropic radial flow and anisotropic
collective flow.

We would like to emphasize that our recent work [31] investigated
particularly the methodology based on the analyses of transverse
momentum spectra, in which the linear relations between $T$ and
$m_0$, $\langle p_T \rangle$ and $m_0$, $\langle p \rangle$ and
$m_0$, $T$ and $\overline{m}$, $\langle p_T \rangle$ and
$\overline{m}$, as well as $\langle p \rangle$ and $\overline{m}$
are studied. At the same time, the relations among some intercepts
and slopes are studied. The present work focuses on an application
of the above methodology. Different kinetic freeze-out
temperatures are obtained due to different types of
``thermometers" or ``thermometric scales" and the same transverse
flow velocity is obtained due to the same transverse momentum
spectrum. Our descriptions on different ``thermometers" or
``thermometric scales" can be compared with those in
thermodynamics.

In addition, the other difference is that in our recent work [31],
the Tsallis distribution and a one- or two-component Erlang
distribution is used, while in the present work, the
Tsallis-standard, Tsallis, and two- or three-component standard
distributions are used. The differences in the description of
particle spectra with the three different distributions are
marginal and only visible in a few cases at high momentum. This is
also reflected in the similar values of the $\chi^2$/dof as listed
in Tables 1 and 2 of the present work. In our opinion, among
different types of ``thermometers" or ``thermometric scales", the
standard distribution and its multi-component distribution can be
used to extract the standard temperature which is closest to the
thermodynamic temperature in thermal physics. The other
``thermometers" can be regarded as non-standard ``thermometers"
which can be corrected if necessary.
\\

{\section{Conclusions}}

We summarize here our main observations and conclusions.

(a) The transverse momentum spectra of final-state light flavour
particles ($\pi^{\pm}$, $K^{\pm}$, $K_S^0$, $p$, $\bar p$, $\phi$,
$\Lambda$, $\bar \Lambda$, $\Xi$, and $\Omega$, etc.), produced in
Au-Au, Cu-Cu, Pb-Pb, $p$-Pb, and $p$-$p$ collisions over an energy
range from 0.2 to 7 TeV, are described by the Tsallis-standard,
Tsallis, and two- or three-component standard distributions which
are used in the multisource thermal model as different types of
``thermometers" or ``thermometric scales" and ``speedometers". The
characteristic of multiple temperatures in multi-component
standard distribution can be described by the Tsallis-standard and
Tsallis distributions. The calculated results are in agreement
with the experimental data recorded by the PHENIX, STAR, ALICE,
and CMS Collaborations. Based on the three types of
``thermometers" or ``thermometric scales" and ``speedometers", the
effective temperatures ($T_{\rm T\textrm{-}S}$, $T_{\rm T}$, and
$T_{\rm S}$) and kinetic freeze-out temperatures ($T_{\rm
T\textrm{-}S0}$, $T_{\rm T0}$, and $T_{\rm S0}$) of interacting
system at the stage of kinetic freeze-out, and the mean transverse
flow velocity of final-state particles are successively extracted
from the transverse momentum spectra.

(b) In the extraction of the kinetic freeze-out temperature and
transverse flow velocity from the transverse momentum spectra, the
methodology used in the present work is different from that of the
original blast-wave model with the Boltzmann-Gibbs distribution or
the Tsallis statistics, or the quark recombination model with the
Cooper-Frye formula and flow. The effect of transverse flow is not
directly considered in formulas describing the transverse momentum
spectra. Instead, we obtain the effective temperature from the
formulas, the kinetic freeze-out temperature from the intercept by
plotting the effective temperature versus the particle's rest
mass, and the transverse flow velocity from the slope by plotting
the mean transverse momentum versus the mean moving mass. Our
treatment disentangles naturally the random thermal motion of
particles and collective expansion of sources.

(c) The present work shows that the effective temperatures $T_{\rm
T\textrm{-}S}$, $T_{\rm T}$, and $T_{\rm S}$ increase with the
increase of centrality or particle's rest mass, and satisfy the
relationship $T_{\rm T\textrm{-}S} \leq T_{\rm T}<T_{\rm S}$ for a
given set of data. In most cases, we have $T_{\rm T\textrm{-}S0}
\leq T_{\rm T0}<T_{\rm S0}$. The kinetic freeze-out temperatures
decrease slightly with the decrease of centrality, and do not show
an obvious change with energy and size for the central,
semi-central, and peripheral collisions. The mean kinetic
freeze-out temperatures derived from the three types of
``thermometers" show $\langle T_{\rm T\textrm{-}S0} \rangle \leq
\langle T_{\rm T0} \rangle < \langle T_{\rm S0} \rangle$, and
$\langle T_{\rm 0} \rangle_{\rm central}
> \langle T_{\rm 0} \rangle_{\rm semi\textrm{-}central} \geq
\langle T_{\rm 0} \rangle_{\rm peripheral}$. The independences of
kinetic freeze-out temperatures on energy and size reveal that
different interacting systems at the stage of kinetic freeze-out
reached the same excitation degree of hadronic matter in the
considered energy range. Generally, the effective temperature is
greater than kinetic freeze-out temperature, since the former
containing both the thermal motion and flow effect while the
latter containing only the thermal motion.

(d) The present work also shows that the central collisions have a
higher kinetic freeze-out temperature than that of peripheral
collisions. This is consistent with those of the chemical
freeze-out temperature and effective temperature. Based on the
three temperatures we conclude that the excitation degree of
central collisions is higher than that of peripheral collisions at
both the chemical and kinetic freeze-out stages. In addition, at
the stage of kinetic freeze-out, the transverse flow velocity in
central collisions is larger than that in peripheral collisions,
indicating a larger squeeze and expansion in central collisions
which also show a higher temperature and excitation degree. Thus,
the kinetic freeze-out temperatures, which describe the random
thermal motions of particles, and the transverse flow velocities,
which describe the collective expansions of sources, paint a
consistent picture.

(e) The present work is an application of our recent work [31] in
which an alternative method is investigated to extract separately
the kinetic freeze-out temperature and flow velocity. If we regard
the standard distribution and its multi-component distribution as
the most exact and standard ``thermometer" due to they being
closest to the thermodynamic temperature in thermal physics, we
obtain a kinetic freeze-out temperature [$(156\pm6)$ MeV in
central collisions] which is equal to the chemical freeze-out
temperature extracted from particle ratios [36]. This indicates
that the two freeze-outs had happened at almost the same time even
at the RHIC and LHC energies. We would like to treat other two
``thermometers" used in the present work as non-standard
``thermometers", which can be corrected if necessary. Another
application [32] of our alternative method shows an evidence of
mass-dependent differential kinetic freeze-out scenario, while the
single and double kinetic freeze-out scenarios are eliminated.

(f) In central collisions, the transverse flow mainly contains the
isotropic radial flow, since the effect of anisotropic collective
flow is small and can be neglected in the extraction of transverse
flow which contains mainly the isotropic radial flow. In contrast,
the transverse flow of peripheral collisions includes mainly the
anisotropic collective flow, since the radial flow is nearly zero.
In this work, the transverse flow contains both the isotropic
radial flow and anisotropic collective flow, which outperforms the
original blast-wave model since it only includes the radial flow.
We also found that the radial flow in central collisions is
greater than anisotropic collective flow in peripheral collisions.
\\

{\bf Acknowledgments}

This work was supported by the National Natural Science Foundation
of China under Grant No. 11575103 and the US DOE under contract
DE-FG02-87ER40331.A008.

\renewcommand{\baselinestretch}{0.8}

\newpage
\begin{sidewaystable}

{\small {Table 1. Values of free parameters, normalization
constants, and $\chi^2$/dof corresponding to Tsallis-standard and
Tsallis distributions in Figs. 1--5. The values of $\chi^2$/dof
for $\Xi^{-}+\bar{\Xi}^{+}$ in Fig. 5(b) are in fact the values of
$\chi^2$ values due to less data points. The temperatures are in
the units of GeV.
{%
\begin{center}
\begin{tabular}{cccccccccc}
\hline\hline  Figure & Type & $T_{\rm T\textrm{-}S}$ & $q_{\rm T\textrm{-}S}$ & $N_{\rm T\textrm{-}S0}$ & $\chi^2$/dof & $T_{\rm T}$ & $q_{\rm T}$& $N_{\rm T0}$ & $\chi^2$/dof \\
\hline
1(a) & 0-5\%                  & $0.115\pm0.003$ & $1.076\pm0.004$ & $184.342\pm28.740$ & 0.243 & $0.120\pm0.003$ & $1.085\pm0.005$ & $169.870\pm25.534$ & 0.311 \\
     & 15-20\%                & $0.113\pm0.003$ & $1.080\pm0.004$ & $109.968\pm16.776$ & 0.193 & $0.119\pm0.003$ & $1.088\pm0.005$ & $102.208\pm17.748$ & 0.314 \\
     & 30-40\%                & $0.108\pm0.003$ & $1.086\pm0.004$ & $54.158\pm8.198$   & 0.347 & $0.112\pm0.003$ & $1.097\pm0.005$ & $52.515\pm7.948$   & 0.377 \\
     & 50-60\%                & $0.096\pm0.003$ & $1.093\pm0.004$ & $21.693\pm3.682$   & 0.273 & $0.103\pm0.003$ & $1.104\pm0.006$ & $20.256\pm3.357$   & 0.324 \\
     & 70-80\%                & $0.086\pm0.003$ & $1.097\pm0.004$ & $5.664\pm1.114$    & 0.234 & $0.094\pm0.003$ & $1.108\pm0.006$ & $5.234\pm1.008$    & 0.278 \\
     & 80-92\%                & $0.078\pm0.003$ & $1.101\pm0.005$ & $3.141\pm0.641$    & 0.149 & $0.087\pm0.003$ & $1.112\pm0.006$ & $2.870\pm0.584$    & 0.182 \\
1(b) & 0-5\%                  & $0.119\pm0.003$ & $1.072\pm0.004$ & $180.218\pm26.650$ & 0.190 & $0.120\pm0.003$ & $1.085\pm0.005$ & $169.269\pm24.783$ & 0.287 \\
     & 15-20\%                & $0.115\pm0.003$ & $1.078\pm0.004$ & $105.190\pm16.405$ & 0.181 & $0.113\pm0.003$ & $1.096\pm0.005$ & $102.669\pm15.936$ & 0.188 \\
     & 30-40\%                & $0.108\pm0.003$ & $1.086\pm0.004$ & $53.048\pm8.028$   & 0.271 & $0.112\pm0.003$ & $1.097\pm0.005$ & $51.420\pm7.818$   & 0.317 \\
     & 50-60\%                & $0.096\pm0.003$ & $1.093\pm0.004$ & $21.259\pm3.550$   & 0.220 & $0.103\pm0.003$ & $1.104\pm0.005$ & $19.834\pm3.249$   & 0.275 \\
     & 70-80\%                & $0.086\pm0.003$ & $1.097\pm0.004$ & $5.542\pm1.013$    & 0.223 & $0.094\pm0.003$ & $1.108\pm0.005$ & $5.120\pm0.965$    & 0.264 \\
     & 80-92\%                & $0.078\pm0.003$ & $1.101\pm0.005$ & $3.093\pm0.641$    & 0.099 & $0.087\pm0.003$ & $1.112\pm0.005$ & $2.826\pm0.584$    & 0.125 \\
1(c) & 0-5\%                  & $0.158\pm0.003$ & $1.066\pm0.005$ & $19.239\pm2.343$   & 0.066 & $0.168\pm0.003$ & $1.071\pm0.006$ & $18.916\pm2.664$   & 0.067 \\
     & 15-20\%                & $0.155\pm0.003$ & $1.067\pm0.004$ & $11.116\pm1.472$   & 0.023 & $0.166\pm0.003$ & $1.072\pm0.006$ & $10.851\pm1.426$   & 0.023 \\
     & 30-40\%                & $0.150\pm0.003$ & $1.068\pm0.004$ & $5.383\pm0.529$    & 0.031 & $0.160\pm0.003$ & $1.073\pm0.006$ & $5.308\pm0.757$    & 0.032 \\
     & 50-60\%                & $0.132\pm0.003$ & $1.074\pm0.005$ & $1.935\pm0.350$    & 0.050 & $0.142\pm0.003$ & $1.080\pm0.006$ & $1.914\pm0.339$    & 0.052 \\
     & 70-80\%                & $0.117\pm0.003$ & $1.080\pm0.006$ & $0.399\pm0.088$    & 0.132 & $0.125\pm0.003$ & $1.088\pm0.006$ & $0.397\pm0.082$    & 0.134 \\
     & 80-92\%                & $0.112\pm0.003$ & $1.082\pm0.006$ & $0.197\pm0.040$    & 0.203 & $0.121\pm0.003$ & $1.090\pm0.006$ & $0.198\pm0.048$    & 0.201 \\
1(d) & 0-5\%                  & $0.158\pm0.003$ & $1.066\pm0.005$ & $18.146\pm2.343$   & 0.040 & $0.168\pm0.003$ & $1.071\pm0.006$ & $17.850\pm2.664$   & 0.034 \\
     & 15-20\%                & $0.155\pm0.003$ & $1.067\pm0.004$ & $10.320\pm1.472$   & 0.041 & $0.166\pm0.003$ & $1.072\pm0.006$ & $10.067\pm1.426$   & 0.036 \\
     & 30-40\%                & $0.150\pm0.003$ & $1.067\pm0.004$ & $5.052\pm0.525$    & 0.030 & $0.161\pm0.003$ & $1.072\pm0.006$ & $4.956\pm0.514$    & 0.029 \\
     & 50-60\%                & $0.132\pm0.003$ & $1.075\pm0.005$ & $1.784\pm0.353$    & 0.059 & $0.144\pm0.003$ & $1.080\pm0.006$ & $1.746\pm0.312$    & 0.062 \\
     & 70-80\%                & $0.112\pm0.003$ & $1.083\pm0.006$ & $0.397\pm0.076$    & 0.085 & $0.121\pm0.003$ & $1.091\pm0.005$ & $0.379\pm0.068$    & 0.085 \\
     & 80-92\%                & $0.108\pm0.003$ & $1.086\pm0.006$ & $0.191\pm0.036$    & 0.178 & $0.116\pm0.003$ & $1.095\pm0.006$ & $0.193\pm0.042$    & 0.174 \\
1(e) & 0-5\%                  & $0.250\pm0.005$ & $1.031\pm0.005$ & $4.940\pm0.820$    & 0.295 & $0.256\pm0.006$ & $1.032\pm0.005$ & $5.082\pm0.847$    & 0.304 \\
     & 15-20\%                & $0.243\pm0.005$ & $1.032\pm0.004$ & $2.929\pm0.460$    & 0.146 & $0.252\pm0.006$ & $1.033\pm0.004$ & $2.943\pm0.475$    & 0.150 \\
     & 30-40\%                & $0.230\pm0.006$ & $1.035\pm0.004$ & $1.493\pm0.302$    & 0.057 & $0.240\pm0.005$ & $1.036\pm0.005$ & $1.492\pm0.314$    & 0.059 \\
     & 50-60\%                & $0.197\pm0.006$ & $1.042\pm0.004$ & $0.559\pm0.107$    & 0.049 & $0.208\pm0.006$ & $1.043\pm0.006$ & $0.555\pm0.104$    & 0.052 \\
     & 70-80\%                & $0.158\pm0.004$ & $1.052\pm0.005$ & $0.126\pm0.032$    & 0.074 & $0.168\pm0.005$ & $1.054\pm0.005$ & $0.127\pm0.027$    & 0.076 \\
     & 80-92\%                & $0.128\pm0.004$ & $1.058\pm0.005$ & $0.070\pm0.023$    & 0.147 & $0.138\pm0.005$ & $1.061\pm0.005$ & $0.069\pm0.025$    & 0.150 \\
1(f) & 0-5\%                  & $0.251\pm0.005$ & $1.029\pm0.004$ & $3.718\pm0.715$    & 0.413 & $0.258\pm0.006$ & $1.030\pm0.005$ & $3.771\pm0.742$    & 0.430 \\
     & 15-20\%                & $0.245\pm0.005$ & $1.030\pm0.004$ & $2.222\pm0.467$    & 0.231 & $0.255\pm0.006$ & $1.030\pm0.004$ & $2.218\pm0.490$    & 0.234 \\
     & 30-40\%                & $0.223\pm0.006$ & $1.034\pm0.004$ & $1.161\pm0.261$    & 0.085 & $0.233\pm0.006$ & $1.035\pm0.005$ & $1.153\pm0.274$    & 0.082 \\
     & 50-60\%                & $0.195\pm0.005$ & $1.041\pm0.004$ & $0.425\pm0.093$    & 0.079 & $0.205\pm0.006$ & $1.042\pm0.006$ & $0.426\pm0.080$    & 0.080 \\
     & 70-80\%                & $0.158\pm0.004$ & $1.049\pm0.005$ & $0.095\pm0.018$    & 0.119 & $0.165\pm0.005$ & $1.052\pm0.005$ & $0.096\pm0.024$    & 0.118 \\
     & 80-92\%                & $0.128\pm0.004$ & $1.055\pm0.005$ & $0.057\pm0.016$    & 0.099 & $0.138\pm0.005$ & $1.057\pm0.005$ & $0.057\pm0.018$    & 0.097 \\
2(a) & 0-10\%                 & $0.123\pm0.007$ & $1.078\pm0.004$ & $7.085\pm2.401$    & 0.243 & $0.132\pm0.007$ & $1.085\pm0.004$ & $7.155\pm2.600$    & 0.242 \\
     & 20-30\%                & $0.118\pm0.007$ & $1.080\pm0.004$ & $3.580\pm1.194$    & 0.289 & $0.128\pm0.007$ & $1.087\pm0.004$ & $3.504\pm1.460$    & 0.290 \\
     & 40-60\%                & $0.115\pm0.007$ & $1.084\pm0.004$ & $1.181\pm0.450$    & 0.196 & $0.125\pm0.007$ & $1.092\pm0.004$ & $1.148\pm0.546$    & 0.195 \\
2(b) & 0-10\%                 & $0.219\pm0.007$ & $1.033\pm0.004$ & $1.624\pm0.722$    & 0.786 & $0.227\pm0.007$ & $1.034\pm0.004$ & $1.620\pm0.733$    & 0.789 \\
     & 20-30\%                & $0.193\pm0.007$ & $1.042\pm0.003$ & $0.778\pm0.303$    & 0.280 & $0.198\pm0.007$ & $1.045\pm0.004$ & $0.801\pm0.280$    & 0.275 \\
     & 40-60\%                & $0.157\pm0.006$ & $1.054\pm0.003$ & $0.278\pm0.104$    & 0.324 & $0.172\pm0.006$ & $1.055\pm0.004$ & $0.270\pm0.104$    & 0.331 \\
2(c) & 0-10\%                 & $0.238\pm0.007$ & $1.031\pm0.003$ & $0.166\pm0.050$    & 0.165 & $0.243\pm0.006$ & $1.033\pm0.003$ & $0.167\pm0.048$    & 0.167 \\
     & 20-30\%                & $0.221\pm0.007$ & $1.039\pm0.003$ & $0.067\pm0.024$    & 0.382 & $0.228\pm0.006$ & $1.042\pm0.004$ & $0.065\pm0.023$    & 0.379 \\
     & 40-60\%                & $0.205\pm0.007$ & $0.040\pm0.003$ & $0.022\pm0.009$    & 0.295 & $0.210\pm0.006$ & $1.043\pm0.003$ & $0.022\pm0.008$    & 0.343 \\
2(d) & 0-10\%                 & $0.299\pm0.007$ & $1.028\pm0.004$ & $0.032\pm0.009$    & 0.856 & $0.306\pm0.007$ & $1.029\pm0.005$ & $0.032\pm0.009$    & 0.858 \\
     & 20-30\%                & $0.227\pm0.007$ & $1.041\pm0.004$ & $0.015\pm0.003$    & 0.092 & $0.233\pm0.007$ & $1.044\pm0.004$ & $0.016\pm0.004$    & 0.089 \\
     & 40-60\%                & $0.215\pm0.007$ & $1.045\pm0.004$ & $0.003\pm0.001$    & 0.081 & $0.223\pm0.007$ & $1.048\pm0.004$ & $0.003\pm0.001$    & 0.080 \\
\hline
\end{tabular}%
\end{center}
}} }
\end{sidewaystable}

\newpage
\begin{sidewaystable}

{\small {Table 1. Continued.
{%
\begin{center}
\begin{tabular}{cccccccccc}
\hline
3(a) & 0-5\%                  & $0.101\pm0.002$ & $1.112\pm0.004$ & $467.262\pm112.560$ & 0.163 & $0.112\pm0.002$ & $1.126\pm0.005$ & $430.964\pm121.590$ & 0.211 \\
     & 50-60\%                & $0.085\pm0.002$ & $1.124\pm0.004$ & $49.274\pm11.070$   & 0.123 & $0.091\pm0.002$ & $1.147\pm0.005$ & $48.274\pm10.284$        & 0.161 \\
     & 80-90\%                & $0.068\pm0.002$ & $1.135\pm0.004$ & $5.142\pm1.318$     & 0.059 & $0.071\pm0.002$ & $1.163\pm0.005$ & $5.183\pm0.952$          & 0.114 \\
3(b) & 0-5\%                  & $0.247\pm0.003$ & $1.040\pm0.004$ & $35.583\pm5.534$    & 0.012 & $0.258\pm0.005$ & $1.038\pm0.005$ & $34.752\pm5.111$         & 0.009 \\
     & 50-60\%                & $0.152\pm0.003$ & $1.092\pm0.004$ & $3.822\pm0.665$     & 0.015 & $0.160\pm0.004$ & $1.107\pm0.005$ & $3.822\pm0.637$          & 0.019 \\
     & 80-90\%                & $0.118\pm0.003$ & $1.106\pm0.004$ & $0.350\pm0.078$     & 0.055 & $0.125\pm0.004$ & $1.123\pm0.005$ & $0.346\pm0.077$          & 0.048 \\
3(c) & 0-5\%                  & $0.368\pm0.005$ & $1.023\pm0.005$ & $7.228\pm1.943$     & 0.319 & $0.380\pm0.010$ & $1.023\pm0.010$ & $7.346\pm1.920$          & 0.362 \\
     & 50-60\%                & $0.290\pm0.005$ & $1.037\pm0.003$ & $0.867\pm0.191$     & 0.038 & $0.305\pm0.006$ & $1.037\pm0.005$ & $0.874\pm0.204$          & 0.041 \\
     & 80-90\%                & $0.195\pm0.005$ & $1.058\pm0.004$ & $0.100\pm0.029$     & 0.122 & $0.200\pm0.005$ & $1.064\pm0.005$ & $0.104\pm0.025$          & 0.104 \\
3(d) & 0-5\%                  & $0.370\pm0.005$ & $1.026\pm0.005$ & $13.520\pm2.914$    & 0.241 & $0.379\pm0.007$ & $1.026\pm0.005$ & $13.487\pm2.810$         & 0.223 \\
     & 50-60\%                & $0.340\pm0.007$ & $1.037\pm0.005$ & $1.419\pm0.229$     & 0.294 & $0.350\pm0.007$ & $1.040\pm0.007$ & $1.400\pm0.262$          & 0.302 \\
     & 80-90\%                & $0.250\pm0.007$ & $1.067\pm0.005$ & $0.094\pm0.026$     & 0.204 & $0.256\pm0.007$ & $1.075\pm0.007$ & $0.096\pm0.021$          & 0.182 \\
4(a) & 0-5\%                  & $0.090\pm0.003$ & $1.137\pm0.004$ & $24.935\pm5.456$    & 0.276 & $0.099\pm0.003$ & $1.163\pm0.005$ & $23.932\pm5.400$         & 0.371 \\
     & 40-60\%                & $0.081\pm0.003$ & $1.137\pm0.004$ & $10.100\pm2.104$    & 0.249 & $0.088\pm0.003$ & $1.163\pm0.006$ & $10.008\pm2.045$         & 0.304 \\
     & 80-100\%               & $0.064\pm0.002$ & $1.139\pm0.004$ & $3.449\pm0.725$     & 0.008 & $0.072\pm0.003$ & $1.164\pm0.006$ & $3.338\pm0.747$          & 0.014 \\
4(b) & 0-5\%                  & $0.273\pm0.008$ & $1.065\pm0.005$ & $0.507\pm0.117$     & 0.107 & $0.295\pm0.009$ & $1.068\pm0.008$ & $0.515\pm0.124$          & 0.102 \\
     & 40-60\%                & $0.174\pm0.005$ & $1.088\pm0.004$ & $0.234\pm0.056$     & 0.026 & $0.184\pm0.006$ & $1.100\pm0.007$ & $0.237\pm0.057$          & 0.032 \\
     & 80-100\%               & $0.082\pm0.004$ & $1.107\pm0.004$ & $0.078\pm0.022$     & 0.019 & $0.091\pm0.004$ & $1.120\pm0.005$ & $0.079\pm0.022$          & 0.018 \\
4(c) & 0-5\%                  & $0.163\pm0.003$ & $1.112\pm0.003$ & $1.901\pm0.242$     & 0.008 & $0.177\pm0.003$ & $1.129\pm0.005$ & $1.915\pm0.231$          & 0.007 \\
     & 40-60\%                & $0.130\pm0.003$ & $1.118\pm0.003$ & $0.739\pm0.098$     & 0.009 & $0.144\pm0.003$ & $1.135\pm0.005$ & $0.750\pm0.094$          & 0.010 \\
     & 80-100\%               & $0.078\pm0.002$ & $1.133\pm0.003$ & $0.234\pm0.028$     & 0.032 & $0.090\pm0.002$ & $1.152\pm0.005$ & $0.235\pm0.033$          & 0.031 \\
4(d) & 0-5\%                  & $0.286\pm0.005$ & $1.066\pm0.003$ & $0.346\pm0.087$     & 0.025 & $0.301\pm0.005$ & $1.072\pm0.004$ & $0.348\pm0.083$          & 0.025 \\
     & 40-60\%                & $0.186\pm0.004$ & $1.087\pm0.003$ & $0.142\pm0.034$     & 0.023 & $0.196\pm0.005$ & $1.098\pm0.004$ & $0.144\pm0.032$          & 0.018 \\
     & 80-100\%               & $0.091\pm0.003$ & $1.106\pm0.003$ & $0.039\pm0.011$     & 0.025 & $0.100\pm0.003$ & $1.119\pm0.003$ & $0.040\pm0.013$          & 0.026 \\
4(e) & 0-5\%                  & $0.165\pm0.004$ & $1.108\pm0.003$ & $0.934\pm0.243$     & 0.002 & $0.179\pm0.004$ & $1.123\pm0.004$ & $0.945\pm0.212$          & 0.005 \\
     & 40-60\%                & $0.130\pm0.003$ & $1.119\pm0.003$ & $0.372\pm0.091$     & 0.016 & $0.145\pm0.004$ & $1.137\pm0.004$ & $0.363\pm0.085$          & 0.026 \\
     & 80-100\%               & $0.080\pm0.003$ & $1.133\pm0.003$ & $0.116\pm0.030$     & 0.015 & $0.091\pm0.003$ & $1.154\pm0.003$ & $0.117\pm0.028$          & 0.016 \\
5(a) & $\pi^+$                & $0.084\pm0.003$ & $1.112\pm0.006$ & $1.862\pm0.317$     & 0.055 & $0.092\pm0.003$ & $1.127\pm0.007$ & $1.852\pm0.308$          & 0.077 \\
     & $K^+$                  & $0.110\pm0.004$ & $1.097\pm0.007$ & $0.230\pm0.029$     & 0.014 & $0.122\pm0.004$ & $1.107\pm0.007$ & $0.229\pm0.029$          & 0.017 \\
     & $p$                    & $0.148\pm0.004$ & $1.065\pm0.005$ & $0.103\pm0.015$     & 0.085 & $0.159\pm0.004$ & $1.069\pm0.007$ & $0.103\pm0.015$         & 0.084 \\
5(b) & $\Lambda$              & $0.166\pm0.003$ & $1.060\pm0.002$ & $0.046\pm0.005$     & 1.189 & $0.175\pm0.003$ & $1.064\pm0.002$ & $0.045\pm0.005$          & 1.192 \\
     & $\phi$                 & $0.164\pm0.003$ & $1.081\pm0.002$ & $0.017\pm0.002$     & 0.861 & $0.174\pm0.003$ & $1.092\pm0.002$ & $0.018\pm0.002$          & 0.804 \\
     & $\Xi^{-}+\bar{\Xi}^{+}$& $0.175\pm0.003$ & $1.055\pm0.002$ & $0.012\pm0.002$     & (0.357) & $0.180\pm0.003$ & $1.060\pm0.004$ & $0.012\pm0.002$        & (0.348) \\
5(c) & $\pi^+$                & $0.084\pm0.003$ & $1.121\pm0.006$ & $2.315\pm0.382$     & 0.106 & $0.092\pm0.003$ & $1.140\pm0.007$ & $2.305\pm0.360$          & 0.131 \\
     & $K^+$                  & $0.112\pm0.003$ & $1.107\pm0.006$ & $0.298\pm0.036$     & 0.009 & $0.123\pm0.003$ & $1.121\pm0.008$ & $0.302\pm0.040$          & 0.014 \\
     & $p$                    & $0.148\pm0.003$ & $1.076\pm0.006$ & $0.133\pm0.019$     & 0.080 & $0.159\pm0.003$ & $1.082\pm0.007$ & $0.134\pm0.021$          & 0.081 \\
5(d) & $\pi^+$                & $0.080\pm0.002$ & $1.130\pm0.004$ & $2.918\pm0.320$     & 0.100 & $0.090\pm0.003$ & $1.150\pm0.005$ & $2.920\pm0.319$          & 0.143 \\
     & $K^+$                  & $0.115\pm0.002$ & $1.110\pm0.004$ & $0.367\pm0.030$     & 0.014 & $0.127\pm0.003$ & $1.124\pm0.005$ & $0.376\pm0.036$          & 0.022 \\
     & $p$                    & $0.148\pm0.002$ & $1.089\pm0.003$ & $0.168\pm0.016$     & 0.044 & $0.163\pm0.003$ & $1.098\pm0.005$ & $0.168\pm0.016$          & 0.042 \\
5(e) & $\Xi$                  & $0.179\pm0.003$ & $1.088\pm0.002$ & $(2.356\pm0.416)\times10^{-3}$ & 0.587 & $0.188\pm0.003$ & $1.099\pm0.002$ & $(2.389\pm0.342)\times10^{-3}$& 0.583 \\
     & $\Omega$               & $0.202\pm0.003$ & $1.086\pm0.002$ & $(2.190\pm0.281)\times10^{-4}$ & 0.045 & $0.215\pm0.003$ & $1.096\pm0.003$ & $(2.180\pm0.322)\times10^{-4}$& 0.037 \\
5(f) & $\pi^{+}+\pi^{-}$      & $0.063\pm0.002$ & $1.148\pm0.004$ & $4.490\pm0.726$     & 0.048 & $0.073\pm0.002$ & $1.173\pm0.005$ & $4.484\pm0.784$          & 0.070 \\
     & $K^{+}+K^{-}$          & $0.095\pm0.002$ & $1.132\pm0.003$ & $0.571\pm0.090$     & 0.017 & $0.107\pm0.003$ & $1.153\pm0.004$ & $0.570\pm0.099$          & 0.021 \\
     & $p+\bar p$             & $0.125\pm0.002$ & $1.097\pm0.002$ & $0.240\pm0.034$     & 0.056 & $0.137\pm0.003$ & $1.108\pm0.003$ & $0.242\pm0.033$          & 0.050 \\
\hline\hline
\end{tabular}%
\end{center}
}} }
\end{sidewaystable}

\newpage
\begin{sidewaystable}

{\small {Table 2. Values of free parameters, $T_{\rm S}$,
normalization constant, and $\chi^2$/dof corresponding to two- or
three-component standard distributions in Figs. 1--5. The values
of $\chi^2$/dof for $\phi$ and $\Xi^{-}+\bar{\Xi}^{+}$ in Fig.
5(b) are the values of $\chi^2$ values due to less data points.
The temperatures are in the units of GeV.
{%
\begin{center}
\begin{tabular}{cccccccccc}
\hline\hline Figure & Type & $T_{1}$ & $k_{1}$ & $T_{2}$ & $k_{2}$ & $T_{3}$ & $T_{\rm S}$ & $N_{\rm S0}$ & $\chi^2$/dof \\
\hline
1(a) & 0-5\%                   & $0.139\pm0.010$ & $0.874\pm0.020$ & $0.276\pm0.010$ & - & - & $0.156\pm0.010$ & $193.000\pm30.000$        & 0.306 \\
     & 15-20\%                 & $0.142\pm0.010$ & $0.896\pm0.020$ & $0.298\pm0.010$ & - & - & $0.158\pm0.010$ & $109.000\pm15.000$        & 0.293 \\
     & 30-40\%                 & $0.142\pm0.010$ & $0.913\pm0.010$ & $0.309\pm0.010$ & - & - & $0.156\pm0.010$ & $54.500\pm8.000$          & 0.457 \\
     & 50-60\%                 & $0.137\pm0.008$ & $0.929\pm0.010$ & $0.315\pm0.010$ & - & - & $0.150\pm0.008$ & $20.500\pm3.000$          & 0.449 \\
     & 70-80\%                 & $0.130\pm0.010$ & $0.935\pm0.010$ & $0.305\pm0.010$ & - & - & $0.141\pm0.010$ & $5.110\pm0.800$           & 0.344 \\
     & 80-92\%                 & $0.121\pm0.010$ & $0.932\pm0.010$ & $0.291\pm0.010$ & - & - & $0.132\pm0.010$ & $2.800\pm0.400$           & 0.237 \\
1(b) & 0-5\%                   & $0.140\pm0.010$ & $0.860\pm0.020$ & $0.271\pm0.010$ & - & - & $0.158\pm0.010$ & $188.000\pm28.000$        & 0.269 \\
     & 15-20\%                 & $0.145\pm0.010$ & $0.891\pm0.020$ & $0.293\pm0.010$ & - & - & $0.161\pm0.010$ & $102.000\pm13.000$        & 0.335 \\
     & 30-40\%                 & $0.143\pm0.008$ & $0.913\pm0.010$ & $0.309\pm0.010$ & - & - & $0.157\pm0.008$ & $52.800\pm7.000$          & 0.483 \\
     & 50-60\%                 & $0.140\pm0.010$ & $0.926\pm0.010$ & $0.315\pm0.010$ & - & - & $0.153\pm0.009$ & $19.320\pm2.000$          & 0.495 \\
     & 70-80\%                 & $0.129\pm0.009$ & $0.926\pm0.010$ & $0.301\pm0.010$ & - & - & $0.142\pm0.008$ & $4.910\pm0.800$           & 0.354 \\
     & 80-92\%                 & $0.122\pm0.010$ & $0.929\pm0.015$ & $0.291\pm0.020$ & - & - & $0.134\pm0.010$ & $2.700\pm0.500$           & 0.245 \\
1(c) & 0-5\%                   & $0.207\pm0.010$ & $0.871\pm0.030$ & $0.390\pm0.030$ & - & - & $0.231\pm0.011$ & $18.400\pm2.000$          & 0.062 \\
     & 15-20\%                 & $0.206\pm0.010$ & $0.866\pm0.030$ & $0.382\pm0.030$ & - & - & $0.230\pm0.011$ & $10.400\pm1.000$          & 0.026 \\
     & 30-40\%                 & $0.202\pm0.010$ & $0.878\pm0.030$ & $0.385\pm0.030$ & - & - & $0.224\pm0.011$ & $5.060\pm0.500$           & 0.040 \\
     & 50-60\%                 & $0.191\pm0.010$ & $0.887\pm0.030$ & $0.362\pm0.030$ & - & - & $0.210\pm0.011$ & $1.810\pm0.200$           & 0.093 \\
     & 70-80\%                 & $0.169\pm0.010$ & $0.845\pm0.030$ & $0.325\pm0.025$ & - & - & $0.193\pm0.010$ & $0.389\pm0.080$           & 0.142 \\
     & 80-92\%                 & $0.165\pm0.015$ & $0.840\pm0.030$ & $0.320\pm0.030$ & - & - & $0.190\pm0.014$ & $0.188\pm0.040$           & 0.200 \\
1(d) & 0-5\%                   & $0.210\pm0.010$ & $0.866\pm0.030$ & $0.384\pm0.030$ & - & - & $0.233\pm0.011$ & $16.800\pm2.000$          & 0.025 \\
     & 15-20\%                 & $0.220\pm0.010$ & $0.912\pm0.030$ & $0.410\pm0.030$ & - & - & $0.237\pm0.011$ & $9.170\pm1.000$           & 0.055 \\
     & 30-40\%                 & $0.202\pm0.010$ & $0.878\pm0.030$ & $0.385\pm0.030$ & - & - & $0.224\pm0.011$ & $4.710\pm0.500$           & 0.038 \\
     & 50-60\%                 & $0.191\pm0.010$ & $0.877\pm0.030$ & $0.362\pm0.020$ & - & - & $0.212\pm0.010$ & $1.650\pm0.200$           & 0.074 \\
     & 70-80\%                 & $0.169\pm0.010$ & $0.855\pm0.030$ & $0.330\pm0.025$ & - & - & $0.192\pm0.010$ & $0.378\pm0.080$           & 0.098 \\
     & 80-92\%                 & $0.165\pm0.015$ & $0.840\pm0.040$ & $0.320\pm0.030$ & - & - & $0.190\pm0.018$ & $0.182\pm0.040$           & 0.184 \\
1(e) & 0-5\%                   & $0.310\pm0.010$ & $0.980\pm0.020$ & $0.520\pm0.030$ & - & - & $0.314\pm0.011$ & $4.500\pm1.000$           & 0.216 \\
     & 15-20\%                 & $0.300\pm0.020$ & $0.900\pm0.030$ & $0.420\pm0.030$ & - & - & $0.312\pm0.019$ & $2.600\pm0.600$           & 0.166 \\
     & 30-40\%                 & $0.264\pm0.020$ & $0.900\pm0.030$ & $0.427\pm0.020$ & - & - & $0.280\pm0.018$ & $1.500\pm0.200$           & 0.061 \\
     & 50-60\%                 & $0.240\pm0.020$ & $0.900\pm0.030$ & $0.400\pm0.030$ & - & - & $0.256\pm0.019$ & $0.560\pm0.100$           & 0.040 \\
     & 70-80\%                 & $0.205\pm0.020$ & $0.920\pm0.030$ & $0.400\pm0.030$ & - & - & $0.221\pm0.019$ & $0.135\pm0.020$           & 0.032 \\
     & 80-92\%                 & $0.180\pm0.030$ & $0.920\pm0.040$ & $0.360\pm0.030$ & - & - & $0.194\pm0.029$ & $0.076\pm0.015$           & 0.117 \\
1(f) & 0-5\%                   & $0.300\pm0.020$ & $0.970\pm0.020$ & $0.490\pm0.030$ & - & - & $0.306\pm0.020$ & $3.500\pm0.800$           & 0.283 \\
     & 15-20\%                 & $0.300\pm0.020$ & $0.970\pm0.020$ & $0.480\pm0.030$ & - & - & $0.305\pm0.020$ & $2.000\pm0.400$           & 0.155 \\
     & 30-40\%                 & $0.265\pm0.020$ & $0.910\pm0.030$ & $0.410\pm0.030$ & - & - & $0.278\pm0.019$ & $1.100\pm0.200$           & 0.070 \\
     & 50-60\%                 & $0.240\pm0.020$ & $0.910\pm0.030$ & $0.400\pm0.030$ & - & - & $0.254\pm0.019$ & $0.430\pm0.100$           & 0.059 \\
     & 70-80\%                 & $0.205\pm0.020$ & $0.890\pm0.030$ & $0.360\pm0.020$ & - & - & $0.222\pm0.018$ & $0.100\pm0.030$           & 0.109 \\
     & 80-92\%                 & $0.180\pm0.020$ & $0.810\pm0.050$ & $0.300\pm0.030$ & - & - & $0.203\pm0.018$ & $0.056\pm0.015$           & 0.104 \\
2(a) & 0-10\%                  & $0.230\pm0.040$ & $0.881\pm0.001$ & $0.350\pm0.040$ & $0.117\pm0.001$ & $0.660\pm0.030$ & $0.245\pm0.036$ & $4.500\pm2.000$ & 0.490 \\
     & 20-30\%                 & $0.217\pm0.050$ & $0.788\pm0.002$ & $0.315\pm0.040$ & $0.208\pm0.001$ & $0.624\pm0.030$ & $0.239\pm0.040$ & $2.100\pm0.900$ & 0.328 \\
     & 40-60\%                 & $0.195\pm0.050$ & $0.792\pm0.001$ & $0.321\pm0.040$ & $0.205\pm0.001$ & $0.650\pm0.030$ & $0.222\pm0.040$ & $0.870\pm0.300$ & 0.206 \\
2(b) & 0-10\%                  & $0.305\pm0.020$ & $0.960\pm0.003$ & $0.360\pm0.030$ & $0.035\pm0.003$ & $0.550\pm0.030$ & $0.308\pm0.021$ & $1.250\pm0.300$ & 0.441 \\
     & 20-30\%                 & $0.270\pm0.020$ & $0.932\pm0.003$ & $0.390\pm0.030$ & $0.060\pm0.003$ & $0.510\pm0.030$ & $0.279\pm0.020$ & $0.700\pm0.200$ & 0.208 \\
     & 40-60\%                 & $0.260\pm0.030$ & $0.950\pm0.003$ & $0.400\pm0.030$ & $0.040\pm0.004$ & $0.500\pm0.030$ & $0.268\pm0.030$ & $0.250\pm0.100$ & 0.350 \\
2(c) & 0-10\%                  & $0.310\pm0.020$ & $0.950\pm0.020$ & $0.465\pm0.020$ & - & - & $0.317\pm0.019$ & $0.150\pm0.050$ & 0.224 \\
     & 20-30\%                 & $0.310\pm0.020$ & $0.950\pm0.020$ & $0.490\pm0.020$ & - & - & $0.319\pm0.019$ & $0.057\pm0.015$ & 0.336 \\
     & 40-60\%                 & $0.295\pm0.020$ & $0.950\pm0.020$ & $0.470\pm0.020$ & - & - & $0.304\pm0.019$ & $0.019\pm0.005$ & 0.324 \\
2(d) & 0-10\%                  & $0.378\pm0.020$ & $0.990\pm0.010$ & $0.610\pm0.050$ & - & - & $0.380\pm0.019$ & $0.030\pm0.010$ & 1.170 \\
     & 20-30\%                 & $0.316\pm0.030$ & $0.940\pm0.020$ & $0.550\pm0.050$ & - & - & $0.330\pm0.029$ & $0.015\pm0.005$ & 0.226 \\
     & 40-60\%                 & $0.308\pm0.030$ & $0.900\pm0.030$ & $0.513\pm0.050$ & - & - & $0.328\pm0.028$ & $0.003\pm0.001$ & 0.203 \\
\hline
\end{tabular}%
\end{center}
}} }
\end{sidewaystable}

\newpage
\begin{sidewaystable}

{\small {Table 2. Continued.
{%
\begin{center}
\begin{tabular}{cccccccccc}
\hline
3(a) & 0-5\%                   & $0.094\pm0.020$ & $0.650\pm0.005$ & $0.254\pm0.020$ & $0.341\pm0.005$ & $0.530\pm0.060$ & $0.152\pm0.030$ & $510.000\pm100.000$       & 0.025 \\
     & 50-60\%                 & $0.083\pm0.020$ & $0.642\pm0.005$ & $0.208\pm0.020$ & $0.325\pm0.005$ & $0.413\pm0.020$ & $0.134\pm0.024$ & $57.000\pm10.000$         & 0.028 \\
     & 80-90\%                 & $0.078\pm0.020$ & $0.635\pm0.005$ & $0.195\pm0.020$ & $0.342\pm0.005$ & $0.430\pm0.020$ & $0.126\pm0.026$ & $5.500\pm1.000$           & 0.021 \\
3(b) & 0-5\%                   & $0.275\pm0.020$ & $0.750\pm0.030$ & $0.389\pm0.020$ & - & - & $0.304\pm0.016$ & $35.000\pm5.000$          & 0.009 \\
     & 50-60\%                 & $0.195\pm0.020$ & $0.680\pm0.040$ & $0.377\pm0.015$ & - & - & $0.253\pm0.016$ & $3.800\pm0.500$           & 0.019 \\
     & 80-90\%                 & $0.160\pm0.020$ & $0.725\pm0.030$ & $0.363\pm0.015$ & - & - & $0.216\pm0.016$ & $0.360\pm0.060$           & 0.048 \\
3(c) & 0-5\%                   & $0.447\pm0.020$ & $0.850\pm0.020$ & $0.300\pm0.020$ & - & - & $0.425\pm0.018$ & $6.800\pm2.000$           & 0.010 \\
     & 50-60\%                 & $0.340\pm0.020$ & $0.950\pm0.010$ & $0.600\pm0.020$ & - & - & $0.353\pm0.019$ & $0.870\pm0.150$           & 0.019 \\
     & 80-90\%                 & $0.228\pm0.020$ & $0.845\pm0.020$ & $0.440\pm0.020$ & - & - & $0.261\pm0.018$ & $0.110\pm0.020$           & 0.065 \\
3(d) & 0-5\%                   & $0.430\pm0.020$ & $0.950\pm0.030$ & $0.670\pm0.050$ & - & - & $0.442\pm0.020$ & $13.500\pm3.000$          & 0.289 \\
     & 50-60\%                 & $0.370\pm0.030$ & $0.850\pm0.030$ & $0.650\pm0.050$ & - & - & $0.412\pm0.028$ & $1.470\pm0.200$           & 0.157 \\
     & 80-90\%                 & $0.290\pm0.020$ & $0.750\pm0.030$ & $0.610\pm0.030$ & - & - & $0.370\pm0.019$ & $0.100\pm0.020$           & 0.060 \\
4(a) & 0-5\%                   & $0.068\pm0.020$ & $0.560\pm0.010$ & $0.200\pm0.020$ & $0.377\pm0.010$ & $0.430\pm0.020$ & $0.140\pm0.014$ & $30.560\pm7.000$          & 0.024 \\
     & 40-60\%                 & $0.071\pm0.020$ & $0.625\pm0.010$ & $0.205\pm0.020$ & $0.335\pm0.010$ & $0.443\pm0.020$ & $0.131\pm0.015$ & $12.000\pm3.000$          & 0.037 \\
     & 80-100\%                & $0.085\pm0.020$ & $0.632\pm0.005$ & $0.195\pm0.015$ & $0.352\pm0.004$ & $0.478\pm0.020$ & $0.130\pm0.030$ & $3.450\pm0.800$           & 0.049 \\
4(b) & 0-5\%                   & $0.283\pm0.050$ & $0.608\pm0.080$ & $0.510\pm0.030$ & - & - & $0.372\pm0.037$ & $0.510\pm0.100$           & 0.117 \\
     & 40-60\%                 & $0.240\pm0.050$ & $0.730\pm0.050$ & $0.480\pm0.030$ & - & - & $0.305\pm0.039$ & $0.230\pm0.050$           & 0.036 \\
     & 80-100\%                & $0.190\pm0.030$ & $0.873\pm0.030$ & $0.445\pm0.030$ & - & - & $0.222\pm0.027$ & $0.080\pm0.020$           & 0.060 \\
4(c) & 0-5\%                   & $0.196\pm0.050$ & $0.556\pm0.080$ & $0.413\pm0.030$ & - & - & $0.292\pm0.035$ & $1.900\pm0.400$           & 0.019 \\
     & 40-60\%                 & $0.175\pm0.050$ & $0.640\pm0.080$ & $0.388\pm0.030$ & - & - & $0.252\pm0.038$ & $0.750\pm0.200$           & 0.018 \\
     & 80-100\%                & $0.150\pm0.040$ & $0.770\pm0.070$ & $0.360\pm0.020$ & - & - & $0.198\pm0.034$ & $0.240\pm0.060$           & 0.042 \\
4(d) & 0-5\%                   & $0.310\pm0.050$ & $0.590\pm0.004$ & $0.520\pm0.030$ & $0.400\pm0.004$ & $0.982\pm0.050$ & $0.401\pm0.032$ & $0.350\pm0.070$           & 0.085 \\
     & 40-60\%                 & $0.250\pm0.050$ & $0.540\pm0.005$ & $0.424\pm0.030$ & $0.442\pm0.005$ & $0.842\pm0.040$ & $0.338\pm0.030$ & $0.140\pm0.030$           & 0.094 \\
     & 80-100\%                & $0.180\pm0.050$ & $0.563\pm0.003$ & $0.358\pm0.030$ & $0.426\pm0.003$ & $0.800\pm0.040$ & $0.263\pm0.031$ & $0.040\pm0.010$           & 0.147 \\
4(e) & 0-5\%                   & $0.210\pm0.050$ & $0.451\pm0.002$ & $0.370\pm0.020$ & $0.530\pm0.002$ & $0.862\pm0.030$ & $0.307\pm0.025$ & $0.940\pm0.200$           & 0.006 \\
     & 40-60\%                 & $0.180\pm0.050$ & $0.576\pm0.002$ & $0.348\pm0.020$ & $0.407\pm0.002$ & $0.840\pm0.030$ & $0.260\pm0.030$ & $0.380\pm0.070$           & 0.006 \\
     & 80-100\%                & $0.130\pm0.040$ & $0.598\pm0.002$ & $0.305\pm0.020$ & $0.390\pm0.002$ & $0.780\pm0.030$ & $0.206\pm0.025$ & $0.120\pm0.020$           & 0.012 \\
5(a) & $\pi^+$                 & $0.124\pm0.010$ & $0.738\pm0.050$ & $0.285\pm0.020$ & - & - & $0.166\pm0.012$ & $1.870\pm0.200$           & 0.069 \\
     & $K^+$                   & $0.165\pm0.010$ & $0.610\pm0.070$ & $0.363\pm0.040$ & - & - & $0.242\pm0.022$ & $0.235\pm0.030$           & 0.030 \\
     & $p$                     & $0.187\pm0.020$ & $0.637\pm0.050$ & $0.380\pm0.040$ & - & - & $0.257\pm0.022$ & $0.106\pm0.010$           & 0.027 \\
5(b) & $\Lambda$               & $0.188\pm0.030$ & $0.610\pm0.060$ & $0.370\pm0.020$ & - & - & $0.260\pm0.023$ & $0.050\pm0.005$           & 1.431 \\
     & $\phi$                  & $0.170\pm0.020$ & $0.600\pm0.060$ & $0.450\pm0.020$ & - & - & $0.282\pm0.022$ & $0.021\pm0.002$           & (0.371) \\
     & $\Xi^{-}+\bar{\Xi}^{+}$ & $0.190\pm0.030$ & $0.730\pm0.050$ & $0.470\pm0.030$ & - & - & $0.266\pm0.030$ & $0.013\pm0.001$           & (0.014) \\
5(c) & $\pi^+$                 & $0.122\pm0.010$ & $0.670\pm0.050$ & $0.285\pm0.020$ & - & - & $0.176\pm0.012$ & $2.300\pm0.300$           & 0.068 \\
     & $K^+$                   & $0.167\pm0.020$ & $0.550\pm0.080$ & $0.368\pm0.040$ & - & - & $0.257\pm0.026$ & $0.305\pm0.040$           & 0.014 \\
     & $p$                     & $0.189\pm0.030$ & $0.555\pm0.100$ & $0.400\pm0.060$ & - & - & $0.283\pm0.038$ & $0.137\pm0.020$           & 0.043 \\
5(d) & $\pi^+$                 & $0.122\pm0.010$ & $0.650\pm0.050$ & $0.296\pm0.020$ & - & - & $0.183\pm0.013$ & $2.900\pm0.300$           & 0.094 \\
     & $K^+$                   & $0.169\pm0.010$ & $0.540\pm0.060$ & $0.368\pm0.040$ & - & - & $0.260\pm0.022$ & $0.375\pm0.030$           & 0.019 \\
     & $p$                     & $0.210\pm0.020$ & $0.510\pm0.090$ & $0.410\pm0.050$ & - & - & $0.308\pm0.032$ & $0.167\pm0.015$           & 0.043 \\
5(e) & $\Xi$                   & $0.270\pm0.040$ & $0.510\pm0.010$ & $0.470\pm0.030$ & $0.450\pm0.010$ & $0.880\pm0.030$ & $0.384\pm0.026$ & $(2.350\pm0.300)\times10^{-3}$ & 0.733 \\
     & $\Omega$                & $0.340\pm0.040$ & $0.680\pm0.050$ & $0.660\pm0.050$ & - & - & $0.442\pm0.035$ & $(2.200\pm0.300)\times10^{-4}$    & 0.084 \\
5(f) & $\pi^{+}+\pi^{-}$       & $0.096\pm0.020$ & $0.584\pm0.010$ & $0.254\pm0.020$ & $0.384\pm0.010$ & $0.650\pm0.040$ & $0.174\pm0.016$ & $4.480\pm0.800$           & 0.061 \\
     & $K^{+}+K^{-}$           & $0.160\pm0.030$ & $0.550\pm0.009$ & $0.358\pm0.020$ & $0.405\pm0.009$ & $0.800\pm0.030$ & $0.269\pm0.020$ & $0.570\pm0.100$           & 0.021 \\
     & $p+\bar p$              & $0.180\pm0.030$ & $0.548\pm0.010$ & $0.360\pm0.020$ & $0.409\pm0.010$ & $0.689\pm0.030$ & $0.276\pm0.019$ & $0.247\pm0.040$           & 0.009 \\
\hline\hline
\end{tabular}%
\end{center}
}} }
\end{sidewaystable}

\end{document}